\def\gsim{~\rlap{$>$}{\lower 1.0ex\hbox{$\sim$}}}
\def\lsim{~\rlap{$<$}{\lower 1.0ex\hbox{$\sim$}}}

\def\wpm2{W m$^{-2}$}

\newcommand{\nc}{\newcommand}
\nc{\teff}{$T_{\rm eff}$\,}
\nc{\logg}{\rm log\,$g$\,}
\nc{\kms}{\,${\rm km\,s}^{-1}$\,}
\nc{\mic}{$\xi_{\rm t}$\,}

\newcounter{foo}
\documentclass[12pt, preprint]{aastex}
\usepackage{CJKutf8}

\begin{document}
\begin{CJK}{UTF8}{bsmi}

\title{Very Low-Mass Stellar and Substellar Companions to Solar-like Stars From MARVELS VI: A Giant Planet and a Brown Dwarf Candidate in a Close Binary System HD~87646}

\author{Bo Ma(馬波)\altaffilmark{1}, 
Jian Ge\altaffilmark{1}, 
Alex Wolszczan$^{2}$, 
Matthew W. Muterspaugh$^{3,4}$, 
Brian Lee$^{1}$,  
Gregory W. Henry$^{4}$, 
Donald P. Schneider$^{2,6}$, 
Eduardo L. Mart\'{\i}n$^7$,  
Andrzej Niedzielski$^{8}$, 
Jiwei Xie$^{1, 9}$,  
Scott W. Fleming$^{14,15}$, 
Neil Thomas$^{1}$, 
Michael Williamson$^{3,4}$, 
Zhaohuan Zhu$^{26}$
Eric Agol\altaffilmark{25}, 
Dmitry Bizyaev$^{10}$, 
Luiz Nicolaci da Costa$^{11,12}$, 
Peng Jiang$^{1,16}$, 
A.F. Martinez Fiorenzano$^{24}$,
Jonay I. Gonz\'alez Hern\'andez$^{22,23}$, 
Pengcheng Guo$^{1}$,
Nolan Grieves \altaffilmark{1}
Rui Li$^{1}$, 
Jane Liu$^{1}$, 
Suvrath  Mahadevan$^{2,6}$, 
Tsevi Mazeh$^{17}$
Duy Cuong Nguyen$^{18}$,
Martin Paegert$^{5}$, 
Sirinrat Sithajan$^{1}$, 
Keivan Stassun$^{5}$, 
Sivarani Thirupathi$^{1}$,
Julian C. van Eyken$^{13}$, 
 Xiaoke Wan$^{1}$, 
Ji Wang$^{19}$,  
John P. Wisniewski$^{21}$, 
Bo Zhao$^{1}$,
Shay Zucker$^{20}$
}
\email{boma@astro.ufl.edu}
\altaffiltext{1}{Department of Astronomy, University of Florida, 211 Bryant Space Science Center, Gainesville, FL, 32611-2055, USA}
\altaffiltext{2}{Department of Astronomy and Astrophysics, The Pennsylvania State University, University Park, PA 16802, USA}
\altaffiltext{3}{Department of Mathematical Sciences, College of Life and Physical Sciences, Tennessee State University, Boswell Science Hall, Nashville, TN 37209, USA}
\altaffiltext{4}{Center of Excellence in Information Systems Engineering and Management, Tennessee State University, 3500 John A. Merritt Blvd., Box No.~9501, Nashville, TN 37209-1561, USA}
\altaffiltext{5}{Department of Physics \& Astronomy, Vanderbilt University, Nashville, TN 37235 USA}
\altaffiltext{6}{Center for Exoplanets and Habitable Worlds, The Pennsylvania State University, University Park, PA 16802, USA}
\altaffiltext{7}{Centro de Astrobiolog\'{\i}a (INTA-CSIC), Carretera de Ajalvir km 4, E-28550 Torrej{\'o}n de Ardoz, Madrid, Spain}
\altaffiltext{8}{Toru{\'n} Centre for Astronomy, Nicolaus Copernicus University in Toru{\'n}, Grudziadzka 5, 87-100 Toru{\'n}, Poland}
\altaffiltext{9}{Department of Astronomy $\&$ Key Laboratory of Modern Astronomy and Astrophysics in Ministry of Education, Nanjing University, Nanjing 210093, China}
\altaffiltext{10}{Apache Point Observatory and New Mexico State University, P.O. Box 59, Sunspot, NM, 88349-0059, USA}
\altaffiltext{11}{Laborat{\'r}io Interinstitucional de e-Astronomia (LIneA), Rio de Janeiro, RJ, 20921-400, Brazil}
\altaffiltext{12}{Observat{\'o}rio Nacional, Rua General Jos{\'e} Cristino, 77, 20921-400 S{\~a}o Crist{\'o}v{\~a}o, Rio de Janeiro, RJ, Brazil}
\altaffiltext{13}{Department of Physics, UC Santa Barbara, Santa Barbara, CA 93106-9530, USA}
\altaffiltext{14}{Space Telescope Science Institute, 3700 San Martin Dr., Baltimore, MD, USA 21218}
\altaffiltext{15}{Computer Sciences Corporation, 3700 San Martin Dr., Baltimore, MD, USA 21218 }
\altaffiltext{16}{Key Laboratory for Research in Galaxies and Cosmology, The University of Science and Technology of China, Hefei, Anhui 230026, China}
\altaffiltext{17}{School of Physics and Astronomy, Raymond and Beverly Sackler Faculty of Exact Sciences, Tel Aviv University, Tel Aviv 69978, Israel}
\altaffiltext{18}{Dunlap Institute for Astronomy and Astrophysics, University of Toronto, Toronto, ON, M5S 3H4, Canada}
\altaffiltext{19}{Department of Astronomy, Yale University, New Haven, CT 06511, USA}
\altaffiltext{20}{Department of Geosciences, Raymond and Beverly Sackler Faculty of Exact Sciences, Tel Aviv University, 6997801 Tel Aviv, Israel}
\altaffiltext{21}{HL Dodge Department of Physics \& Astronomy, University of Oklahoma, 440 W Brooks St, Norman, OK 73019, USA}
\altaffiltext{22}{Instituto de Astrof{\'\i}sica de Canarias, E-38205 La Laguna, Tenerife, Spain}
\altaffiltext{23}{Universidad de La Laguna, Dpto. Astrof{\'\i}sica, E-38206 La Laguna, Tenerife, Spain}
\altaffiltext{24}{Fundaci{\'o}n Galileo Galilei-INAF, Rambla Jos{\'e} Ana Fernandez P{\'e}rez, 738712 Bre{\~n}a Baja, Tenerife, Spain}
\altaffiltext{25}{Department of Astronomy, University of Washington, Box 351580, Seattle, WA 98195-1580, USA}
\altaffiltext{26}{Department of Astrophysics, Princeton University, Princeton, NJ, 08544, USA}

\begin{abstract}
We report the detections of a giant planet (MARVELS-7b) and a brown dwarf  
candidate (MARVELS-7c) around the primary star in the close binary system, HD~87646. 
It is the first close binary system with more than one substellar
circum-primary companion discovered to the best of our knowledge.
The detection of this giant planet was accomplished using the first 
multi-object Doppler instrument (KeckET) at the Sloan Digital Sky Survey (SDSS) telescope. 
Subsequent radial velocity observations using ET at Kitt Peak National Observatory, HRS at HET, 
the ``Classic" spectrograph at the Automatic Spectroscopic Telescope at Fairborn Observatory, 
and MARVELS from SDSS-III confirmed this giant planet 
discovery and revealed the existence of a long-period brown dwarf in this binary. 
HD~87646 is a close binary with a separation of $\sim22$ AU between 
the two stars, estimated using the Hipparcos catalogue and our newly acquired AO image 
from PALAO on the 200-inch Hale Telescope at Palomar. 
The primary star in the binary, HD~87646A, has \teff = 5770$\pm$~80K, \logg=4.1$\pm$0.1 
and [Fe/H] = $-0.17\pm0.08$. The derived minimum masses of the two substellar 
companions of HD~87646A are 12.4$\pm$0.7 M$_{\rm Jup}$ 
and 57.0$\pm3.7$ M$_{\rm Jup}$. The periods are 13.481$\pm$0.001 days 
and 674$\pm$4 days and the measured eccentricities are 0.05$\pm$0.02 
and 0.50$\pm$0.02 respectively. Our dynamical simulations show the system is stable 
if the binary orbit has a large semi-major axis and a low eccentricity, which can 
be verified with future astrometry observations. 
\end{abstract}

\section{INTRODUCTION} 
One of the most surprising astronomical developments of the last 25 years has been 
the discovery of an abundant population of extra-solar planets and brown dwarfs (BDs) 
\citep{zan92, mayor95, rebolo95, nakajima95}. 
Radial velocity (RV) surveys to date have detected over 500 extrasolar planets \citep{han14}.  
The classical planet formation paradigm, that giant planets form and reside only 
in circular orbits at large distances from 
their parent stars, works well for our solar system, but not for extrasolar 
planetary systems. These RV extrasolar planets reveal an astonishing diversity of 
masses, semi-major axes and eccentricities, from the short period hot Jupiters, to planets 
in very elongated orbits, to planetary systems with multiple Jupiter-mass planets, to the 
super-Earth-mass planets with orbital periods of a few days (Butler et al. 2004; 
McArthur et al. 2004; Santos et al. 2004; Rivera et al. 2005; Lovis et al. 2006; 
Udry et al. 2006; Howard et al. 2010; Howard et al. 2014). Indeed, if any single statement 
captures the developments of this field, it is that the observations have continually revealed an 
unanticipated diversity of planetary systems.

Despite the fact that over 500 known exoplanets have provided important information 
about planet masses and orbital parameters, many more exoplanets are urgently 
needed for statistical characterization of emerging classes of planets and tests of detailed 
theoretical models for planet formation and evolution. A large planet sample is also
needed to study the correlation between the presence of planets and stellar 
properties, such as metallicity, mass, multiplicity, age, evolutionary stage, activity level, 
and rotation velocity, not to mention the discovery of new planet populations. 
The growing need of more exoplanet samples triggered the development 
of multi-object Doppler technology at the University of Florida in 2004-2008. 
The first full-scale, multi-object Exoplanet Tracker instrument, the W.M. Keck Exoplanet Tracker (KeckET), was constructed in August 2005-February 2006 with 
Keck Foundation support. It was coupled to a wide field Sloan Digital 
Sky Survey telescope (SDSS, Gunn et al. 2006) and used for the pilot 
Multi-Object APO Radial Velocity Exoplanet Large-Area Survey (MARVELS; Ge et al. 2008, 2009; 
Ge \& Eisenstein 2009) in 2006-2007 \citep{fleming10, eisenstein11}. 

This is the sixth paper in this series examining the low-mass 
companions around solar type stars from the SDSS-III MARVELS survey (Wisniewski et al. 2012; 
Fleming et al. 2012; Ma et al. 2013; Jiang et al. 2013; De Lee et al. 2013). 
In this paper, we present our discovery of two substellar 
companions, a giant planet (MARVELS-7b) and a brown dwarf (MARVELS-7c), 
around the primary star in a binary system, HD~87646, from the MARVELS pilot 
planet survey program. 
Section 2 reviews our previous knowledge about HD~87646. 
Section 3 introduces a brief description of the multi-object Doppler instrument  and the pilot survey. 
Section 4 summarizes the survey data processing pipeline, 
Section 5 describes additional observations of the system, and 
Section 6 gives details of the results. Section 7 presents the main results and a
discussion. 

\section{HD~87646 \label{sec:star}}
The target star, HD~87646, is a bright (V=8) G-type star with a fainter K-type stellar 
companion at a separation of  0.213 arcseconds and a position angle (PA) of 
136 degrees according to the Hipparcos and Tycho Catalogues \citep{perryman97}. 
Its Hipparcos parallax of $13.59 \pm 1.58$ milliarcsecond (mas) places 
it at a distance of $73.58 \pm 9.68$ pc. Photometry and high resolution spectroscopic observation of 
HD~87646A have been obtained by \citet{feltzing98}. They obtained a effective 
temperature of \teff$=5961$~K from photometry, and spectroscopically 
derived $\logg=4.41$  and $\rm [Fe/H] = 0.3$. 
This star is quite metal rich, prompting \citet{Gonzalez01} to explicitly recommend 
that it be observed with precise radial velocity instruments due to the significantly 
higher probability of discovering hot Jupiter planets around metal rich stars 
\citep{Fischer05}. While HD~87646 was observed as part of the Geneva 
Copenhagen Survey \citep{Nordstrom08} we are presently unaware of 
any high precision radial velocity observation, or the star being part of any 
ongoing surveys.

Such binary systems are challenging for precision radial velocity detection due 
to the complexity of analyzing spectra from two different stars. 
While detections of exoplanets in unresolved stellar binaries have 
been reported before \citep{Konacki05}, higher precision observations 
and better cadence have not detected the same signal \citep{Eggenberger07}. 
We speculate that this difficulty in detection is the reason this target was not 
observed in some ongoing surveys, like N2K \citep{Robinson06}, that 
target high-metallicity stars. Multi-object surveys do not need to be as 
selective due to their inherent multiplicity advantage. Binaries may be 
excluded, but the existence of a few binaries among 60 stars observed 
simultaneously is not a significant problem. In addition, once observations 
are well underway there is little advantage gained in removing the 
target since any replacement target would then only be observed for a few epochs. 
We will study the impact of spectral contamination from a faint companion star 
on the RV measurements for our target in Section~\ref{sec:contamination}. 
Our study shows that the only substellar companions that can be detected in such close 
binaries are those massive enough to generate RV signals much larger than the noise 
induced by the spectral contamination.

\section{The Multi-object KeckET pilot survey}

The design of KeckET is based on a single object Exoplanet Tracker (ET) 
design for the KPNO 2.1m telescope (Ge et al. 2003, 2006; 
Mahadevan et al. 2008). This instrument adopts the dispersed 
fixed-delay interferometry (DFDI) approach for Doppler measurements 
(Erskine \& Ge 2000; Ge 2002; Ge et al. 2002). Instead of the line centroid shifts 
in the high resolution cross-dispersed echelle spectrograph approach, the 
DFDI method measures the Doppler motion by monitoring the fringe shifts 
of stellar absorption lines created by a Michelson-type interferometer 
with a fixed-delay between the two interferometer arms. The measurement of
this fixed-delay is described in \citet{wang12a,wang13}.

The KeckET instrument consists of 8 subsystems--a multi-object fiber feed, 
an iodine cell, a fixed-delay interferometer system, a slit, a collimator, 
a grating, a camera, and a 4k$\times$4k CCD detector. In addition, 
it  contains four auxiliary subsystems: the interferometer
control, an instrument calibration system, a photon flux monitoring 
system, and a thermal probe and control system. The instrument is 
fed with 60 fibers with 200 $\mu$m core diameters, which are coupled 
to 180 $\mu$m core diameter short fibers from the SDSS telescope, 
corresponding to 3 arcsec on the sky at $f/5$ (Ge et al. 2006c). The 
resolving power for the spectrograph is R=5100, and the wavelength 
coverage is $\sim$ 900~\AA, centered at 5400~\AA. Details of the 
instrument design can be found in Ge et al. (2006b), Wan et al. (2006), 
and Zhao \& Ge (2006). KeckET has one spectrograph and one 
4k$\times$4k CCD camera that captures one of the two interferometer 
outputs, and has a 5.5\% detection efficiency from the telescope to the 
detector without the iodine cell under the typical APO seeing conditions 
($\sim$1.5 arcsec seeing). The CCD camera  records fringing spectra 
from 59 objects in a single exposure.

KeckET was commissioned at the SDSS telescope in Spring 2006. 
After a few engineering upgrades in Fall 2006, we conducted a pilot 
planet survey of 700 FGK main sequence stars in 12 fields 
with V = 7.6-12 to detect new planets in December 2006 to May 2007. 
A total of 5-25 RV measurements have been obtained for the survey 
stars. The data were processed by a modified version 
of data pipeline for the KPNO ET (Ge et al. 2006; van Eyken 2008). 
The instrument Doppler precision was measured 
with the day sky scattered light, which offers a stable, homogeneous 
RV source for simultaneously calibrating the instrument performance 
for all of the sky fibers. The rms error averaged over the 59 fibers, 
measured from the dispersion of measurements over a several hour 
interval in 2006 November, is 6.3$\pm$1.3~${\rm m \, s^{-1}}$. The corresponding 
average photon-limit error is 5.5$\pm$0.5~${\rm m \, s^{-1}}$. The instrument's 
precision over longer time intervals has been measured with repeated 
observations of sky scattered light over a period of 45 days in Fall 2006, 
and 150 days in Winter/Spring 2007. The rms dispersion of 
RV measurements of the sky over these periods, after subtracting the photon 
limiting errors in quadrature, are 11.7$\pm$2.7~${\rm m \, s^{-1}}$ 
and 11.3$\pm$2.5~${\rm m \, s^{-1}}$, respectively. 

The instrumental contributions to random measurement errors are 
mainly caused by inhomogeneous illumination of the slit, image 
aberration, and the interferometer comb aliasing (sampling on the detector). 
However, the dominant measurement RV error is produced by the mathematical approximation 
used for extracting iodine and stellar Doppler signals in the mixed stellar 
and iodine fringing spectra, which is on the order of 50~${\rm m \, s^{-1}}$ 
(van Eyken, Ge \& Mahadevan 2010), which is included in the RV errors showing in 
the data table.  Although this error has largely limited 
our capability of detecting relatively low mass planets, it does not affect the 
Doppler detection of massive giant planets, brown dwarfs and binaries.

\section{Survey Data Processing and RV Results}

The pipeline processing steps are described in detail in van 
Eyken et al. (2004), Ge et al. (2006), Mahadevan et al. (2008) 
and van Eyken et al. (2010). The data were processed using standard 
IRAF procedures \citep{tody93}, as well as software written in IDL. 
The images were corrected for biases, dark current, and scattered 
light and then trimmed, illumination corrected, slant corrected and low-pass 
filtered. The visibilities ($V$) and the phases 
($\theta$) of the fringes were determined for each channel by fitting a sine 
wave to each column of pixels in the slit direction. To determine differential 
velocity shifts the star+iodine data can be considered as a summation of the 
complex visibilities (${\bf V}=Ve^{i\theta}$) of the relevant star 
($V_Se^{i\theta_{S_0}}$) and iodine ($V_Ie^{i\theta_{I_0}}$) templates 
( Erskine 2003; van Eyken et al. 2004, 2010). For small 
velocity shifts the complex visibility of the data, for each wavelength 
channel, can be written as
\begin{equation}
\label{eq:phase}
V_D e^{i\theta_D} = V_S
e^{i\theta_{S_0}}e^{i\theta_S-i\theta_{S_0}}
+V_Ie^{i\theta_{I_0}}e^{i\theta_I-i\theta_{I_0}},
\end{equation}
where $V_D$, $V_S,$ and $V_I$ are the fringe visibilities for a given 
wavelength in the star+iodine data, star template, and iodine template 
respectively, and $\theta_D$, $\theta_{S_0}$, and $\theta_{I_0}$ are the 
corresponding measured phases. In the presence of velocity shifts of 
the star and instrument drift, the complex visibilities of the star and iodine 
template best match the data with phase shifts of $\theta_S - \theta_{S_0}$ 
and $\theta_I - \theta_{I_0}$, respectively. The iodine is a stable reference 
and the iodine phase shift tracks the instrument drift. The difference 
between star and iodine shifts is the real phase shift of the star, $\Delta \phi$, 
corrected for any instrumental drifts
\begin{equation}
\Delta \phi = (\theta_S - \theta_{S_0}) - (\theta_I -
\theta_{I_0}).
\end{equation}
This phase shift can be converted to a velocity shift, $\Delta v$, using a 
known phase-to-velocity scaling factor: 
\begin{equation}
\label{eq:rv}
\Delta v = \frac{c \lambda}{2\pi d}  \Delta \phi,
\end{equation}
where $c$ is the speed of light, $\lambda$ is the wavelength and $d$ is the 
optical delay in the Michelson interferometer. 
The KeckET data analysis pipeline identifies the shift in phase of the star and iodine 
templates that are the best match for the data, and uses these phase shifts to
calculate the velocity shift of the star relative to the stellar template. 
Since HD~87646 is a close binary system, we need an additional 
complex visibility term in equation~\ref{eq:phase} to account for the contamination of the secondary 
star. Mathematically this is equivalent to adding a small noise term $\delta\theta_S$ to the phase of the 
primary star $\theta_S$ \citep{vaneyken10}, which will translate to the measured RV according to 
equation~\ref{eq:rv}. This noise term depends mainly on the flux ratio of 
the two stars collected through the fiber and the radial velocity offset between the two stars. 
Thus this noise term varies slowly between observations. To simplify the RV fitting process, 
we treat this noise as a constant value and will study its impact 
on the RV measurements of HD~87646 in section~\ref{sec:contamination}. In practice, 
one can try to model the visibility and phase of the secondary star if both star spectra 
and their flux ratio variations with wavelength are known precisely. 
This method is not very practical in our current case because of the lack of these information.

We have obtained a total of 16 observations of HD~87646 using KeckET 
from 2006 December to 2007 June. The 
radial velocities obtained are listed in Table \ref{tbl:KeckETvels}.

\begin{table}
\caption{KeckET Pilot Project Radial Velocities for HD~87646. A total of 16 
observations are listed here. }
\begin{tabular}{ c c c}
    \hline
    Julian Date (UTC) & Velocity & Velocity Error \\ 
         & m~s$^{-1}$ & m~s$^{-1}$ \\ 
    \hline
        2454101.86236   &     21983    &    52 \\
        2454102.02955   &     22070    &    54\\
        2454105.97151   &     21908    &    53 \\
        2454128.81306   &     21945    &    52 \\
        2454136.77204   &     20754    &    52 \\
        2454136.80788   &     20753    &    51 \\
        2454164.75013   &     20946    &    53 \\
        2454165.74965   &     20977    &    52 \\
        2454165.78557   &     20935    &    52 \\
        2454186.65127   &     22231   &     51 \\
        2454191.72256   &     21287    &    53 \\
        2454194.73526   &     21359    &    53 \\
        2454195.72733   &     21850    &    52 \\
        2454221.62148   &     21270    &    51 \\
        2454224.61552   &     23040    &    52 \\
        2454254.63197   &     21925    &    55 \\
    \hline
    \label{tbl:KeckETvels}
\end{tabular}
\end{table}

\section{Follow-up Observations}

\subsection{KPNO ET RV Observations}
Subsequent observations were performed using the Exoplanet Tracker (ET) instrument at 
KPNO \citep{ge06c}. Initial follow-up was performed in November of 2007, which  
confirmed the variability seen in the KeckET data. Additional data points were obtained
at KPNO in 2008 January, 2008 February and 2008 May.  The integration time 
was 35-40 mins in 2007 November and 20 mins in 2008 January, 2008 
February and 2008 May.

The data were reduced using software described in \citet{mahadevan08} and references therein.  
See \citet{vaneyken10} for the theory behind the technique.  A total of 40 data points 
were obtained from 2007 November to 2008 May and are listed in Table~\ref{tbl:KPNOvels}.  
The observations confirmed the linear trend shown in the KeckET data, which is later 
to be found due to another substellar companion.

\begin{table}
  \caption{KPNO ET Radial Velocities for HD~87646. This table is 
  available in its entirety in the online journal. There are a total 
  of 40 observations.  
  A portion is shown here for guidance regarding its form and content.}
  \begin{tabular}{ c c c}
    \hline
    Julian Date (UTC) & Velocity & Velocity Error \\ 
         & m~s$^{-1}$ & m~s$^{-1}$ \\ 
          \hline
        2454425.90567    &    19968   &     51 \\
        2454425.98306    &    19869   &     51 \\
        2454427.90591    &    19207   &     52 \\
        2454427.98285    &    19219   &     53 \\
        2454428.98821    &    19896   &     52 \\
        2454430.00529    &    20101  &     54 \\
        2454431.90248    &    19143   &     54 \\
        2454431.97582    &    19159   &     55 \\
        2454491.88155    &    20158   &     55 \\
        2454491.93285    &    20230   &     54 \\
    \hline
    \label{tbl:KPNOvels}
\end{tabular}
\end{table}

\subsection{HET RV observations}
 Follow-up observations of HD~87646 were conducted with the fiber-fed High Resolution 
Spectrograph (HRS, Tull 1998) of the Hobby Eberley telescope \citep[HET,][]{ramsey98}. 
The observations 
were executed in queue scheduled mode (Shetrone et al. 2007), and used a 2 arcsecond 
fiber and with the HRS slit set to yield a spectral resolution of $R\sim60,000$. A total of 
29 data points were obtained between 2007 December and 2008 March.
An iodine cell was inserted into the beam path to yield a fiducial velocity reference. 
The radial velocities were obtained using the procedure and analysis techniques described 
in \citet{marcy92}. The HRS spectra consisted of 46 echelle orders recorded on the 
blue CCD (407$-$592 nm) and 24 orders on the red one (602$-$784 nm). The spectral
 data used for RV measurements were extracted from the 17 orders (505$-$592 nm) in 
which the I$_2$ cell superimposed strong absorption lines. The radial velocities 
obtained are listed in Table~\ref{tbl:HETvels}.

\begin{table}
  \caption{HET Radial Velocities for HD~87646. This table is 
  available in its entirety in the online journal. There are a total 
  of 29 observations.  
  A portion is shown here for guidance regarding its form and content.}
  \begin{tabular}{ c r c}
    \hline
    Julian Date (UTC) & Velocity & Velocity Error \\ 
         & m~s$^{-1}$ & m~s$^{-1}$ \\ 
    \hline
2454454.85471  &   -192.5 & 44.0 \\
2454454.85762  &   -201.0 & 34.3 \\
2454455.86358  &    198.4 & 35.6 \\
2454455.86528  &    192.1 & 31.8 \\
2454458.86351  &    -746.7 & 12.9 \\
2454461.84553  &   -1508.8 & 37.3 \\
2454462.82117  &   -1191.1 & 18.6 \\
2454463.83110  &    -678.3 & 12.3 \\
2454470.81807  &     144.1 & 21.8 \\
2454473.78993  &   -1332.5 & 17.0 \\
    \hline
    \label{tbl:HETvels}
  \end{tabular}
\end{table}

\subsection{MARVELS RV observations}
HD~87646 was selected as an RV survey target by the MARVELS 
preselection criterion (Paegert et al. 2015).  
The star has been monitored at 23 epochs using the 
MARVELS instrument mounted on the SDSS 2.5m Telescope at APO between 
2009 May and 2011 December (Ge et al. 2008, Ma et al. 2013). The MARVELS 
instrument is a fiber-fed dispersed fixed-delay interferometer instrument 
capable of observing 60 objects simultaneously and covers a wavelength range of 5000$-$5700~\AA $\;$ 
with a resolution of R$\sim$12,000. The data processing and error estimation algorithm have been 
described in detail by \citet{Thomas16}. The final differential RV products are 
included in the SDSS Data Release 12 \citep{alam15} and are presented in Table~\ref{tbl:MARVELS}.

\begin{table}
  \caption{MARVELS Radial Velocities for HD~87646. This table is 
  available in its entirety in the online journal. There are a total 
  of 23 observations.  
  A portion is shown here for guidance regarding its form and content.}
  \begin{tabular}{ c r c}
    \hline
    Julian Date (UTC) & Velocity & Velocity Error \\ 
         & m~s$^{-1}$ & m~s$^{-1}$ \\ 
    \hline
       2454960.63450  &         121  &        41 \\
       2454961.63480  &          33  &        41 \\
       2455193.99281  &        -652  &        51 \\
       2455197.91351  &        -529  &        43 \\
       2455223.8140  &        -186  &        46 \\
       2455284.82871  &       -1785  &        53 \\
           \hline
    \label{tbl:MARVELS}
  \end{tabular}
\end{table}

\subsection{Fairborn RV observations}

To investigate the nature of the linear RV trend found in previous RV data, we have obtained 
additional observations of HD~87646 with a fiber-fed echelle spectrograph 
situated at the 2~m Automatic Spectroscopic Telescope (AST) in 
Fairborn Observatory (Eaton \& Williamson 2004, 2007). The robotic nature 
of the AST allowed for high cadence observations, which removed orbital 
period degeneracies and helped solidify the longer-period companion's orbit.
Through 2011 June the detector was a 2048$\times$4096 SITe ST-002A 
CCD with 15 micron pixels. The AST echelle spectrograph has 
21 orders that cover the wavelength range 4920$-$7100~\AA, 
and has an average resolution of 0.17~\AA. 
Beginning in 2010 January, several upgrades were made
to increase the throughput, sensitivity and flexibility of the AST \citep{fekel13}. 
In the summer of 2011 the SITe CCD detector and dewar were replaced
with a Fairchild 486 CCD having 4K$\times$4K 15 micron pixels, which required 
a new readout electronics package, and a new dewar with a 
Cryotiger refrigeration system. The echelle spectrograms that 
were obtained with this new detector have 48 orders, covering 
the wavelength range 3800$-$8260~\AA. The data reduction and radial 
velocity measurements are discussed in Eaton \& Williamson (2007). 
A total of 135 data points were obtained from 2009 March through 
2013 October and are listed in Table~\ref{tbl:Fairbornvels}. With these 
additional RV data, we were able to detect the turnaround of the previous 
identified linear RV trend and start to uncover the second substellar 
companion's orbit.

\begin{table}
  \caption{Fairborn Radial Velocities for HD~87646. This table is 
  available in its entirety in the online journal. There are a total 
  of 135 observations.  
  A portion is shown here for guidance regarding its form and content.}
  \begin{tabular}{ c r c}
    \hline
    Julian Date (UTC) & Velocity (ms$^{-1}$) & Velocity Error (ms$^{-1}$) \\
    \hline
         2454903.7521   &       21690   &    212   \\
         2454904.8249   &       20810   &    289   \\
         2454905.6178   &       21580   &    187   \\
         2454906.7753   &       21470   &    258   \\
         2454908.7537   &       21290   &    142   \\
         2454909.7757   &       21530   &    179   \\
         2454910.7670   &       21600   &    342   \\
         2454911.7909   &       22720   &    171   \\
         2454913.7546   &       22740   &    214   \\
         2454914.7619   &       22140   &    217   \\
    \hline
    \label{tbl:Fairbornvels}
  \end{tabular}
\end{table}

\subsection{TNG high resolution spectroscopy}
A total of nine high-resolution spectra ($R=164,000$) of HD~87646 
were obtained with the SARG spectrograph on the 3.5-m TNG telescope 
at La Palma on 2008 March 21, 22, 28, April 03 and April 11. These data 
were used to monitor line bisector variations, to determine stellar 
properties (metallicity, $\logg$, T$_{\rm eff}$ and $v \sin i$), and to search
for evidence of a second set of lines in the system.
The typical signal-to-noise ratio (S/N) for each spectrum 
is about 150 per resolution element around $\rm 5500\AA$.

\subsection{KPNO EXPERT high resolution spectroscopy}
We also obtained spectroscopic observations of HD~87646 
from the 2.1m telescope at Kitt Peak National Observatory using the $R=30,000$ 
Direct Echelle Mode of the EXPERT spectrograph (Ge et al. 2010). A total of 
seven EXPERT spectra were acquired between 2014 Feb to 2014 June. The 
exposure time for each observation ranged from 20-40 minutes, yielding an 
$S/N \sim 250$ per resolution element around 5500\AA.

\subsection{Lucky imaging}
\begin{figure}
\vspace{-0.35cm}
\centerline{\includegraphics[width=0.7\textwidth]{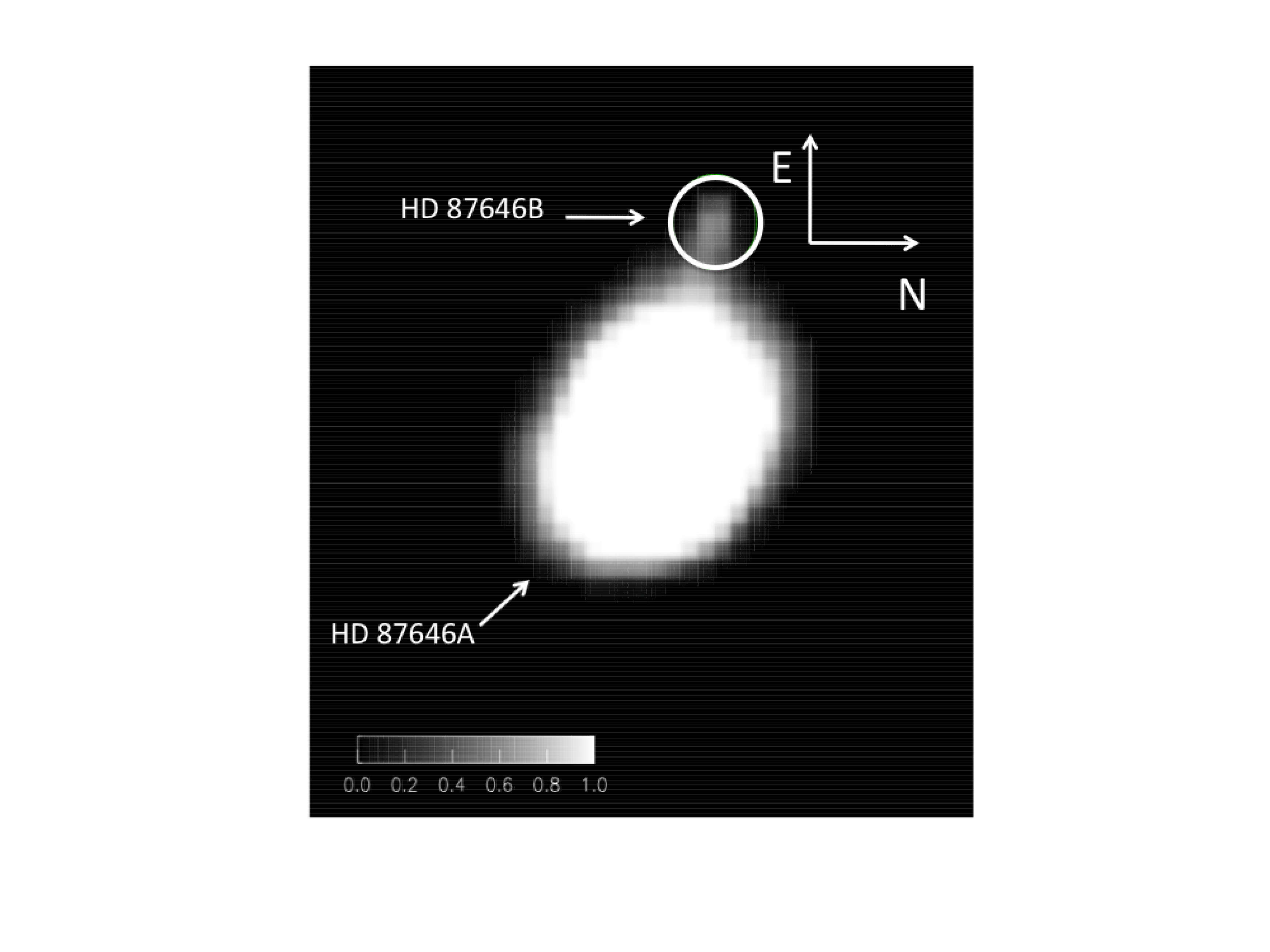}}
\vspace{-0.5in}
\caption{A Lucky image of the HD~87646 system, generated 
by processing and co-adding only the best $1\%$ of the images taken by FastCam on 
the 2.5-meter Nordic Optical Telescope at the Roque de los Muchachos Observatory 
in La Palma, Spain. The secondary star, HD~87646B, is highlighted by the solid white 
circle. The image has a scale of 31~mas~pixel$^{-1}$ and the star PSF has a FWHM of $0.11"$. }
\label{fig:luckyimage}
\end{figure}

On 2008 May 29, high angular resolution lucky images of HD~87646 were 
obtained with the FastCam instrument \citep{oscoz08}
on the 2.5-meter Nordic Optical Telescope at the Roque de los 
Muchachos Observatory in La Palma (Spain).
Five data cubes of 1000 images each were obtained using the I-band filter.
Individual exposure times were 30 ms for each image.
High spatial resolution was obtained by combining the best 1\% of
the images. Fig.~\ref{fig:luckyimage} shows the processed image. The image scale was 30.95 $\pm$ 0.05 mas~pixel$^{-1}$. 
In this figure the secondary star HD~87646B in this binary system is visible. The point spread function (PSF) of the star
has a full width half maximum (FWHM) value of $0.11"$.

\subsection{Palomar AO imaging}

On 2009 June 4, we acquired high resolution AO images of the binary
star system HD~87646 from PALAO on the 200-inch Hale Telescope at 
Palomar. Data sets were taken in the J and K bands.  
The AO system was running at 500 Hz. The seeing was roughly 1.3" in 
K band.  We also observed a calibrator star and subtracted the K-band's PSF, 
which improved sensitivity by a factor of a few. The utility of PSF subtraction 
is limited, in this case, by the difference in stellar spectral types, since the filters 
are broad.

\begin{figure}
\vspace{-0.1cm}
\centerline{\includegraphics[width=0.5\textwidth]{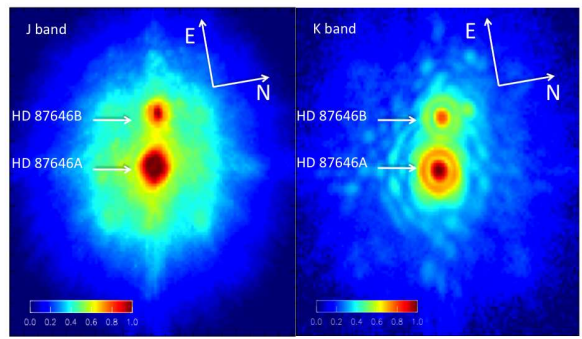}}
\vspace{-0.5in}
\caption{J and K band AO imaging of HD~87646 taken at Palomar observatory. 
The images are displayed on a logarithmic scale, which show the two stars have a 
separation of $401\pm12$ mas. The images have a scale of 25 mas~pixel$^{-1}$. The FWHMs are 
70~mas and 120~mas for J and K band images respectively. }
\label{fig:AO}
\end{figure}

Images were flat-fielded, background subtracted and cleaned; the final images are displayed on 
a logarithmic scale in Fig.~\ref{fig:AO} with a scale of 25 mas~pixel$^{-1}$. The binary system is well-resolved in both bands. The angular 
separation is measured to be $401\pm12$ mas, nearly twice that quoted from
the Hipparcos and Tycho Catalogues \citep{perryman97}. The position angle is $69.8^{\circ}\pm0.5^{\circ}$. 
There is no evidence in the high resolution images for a tertiary (stellar) companion. The J and K-band 
brightness ratios of the two stars are $6.18\pm0.12$ and $5.82\pm0.10$, respectively. We can 
not use these ratios and their error bars to put a meaningful constraint on the optical band flux ratio 
because the spectral energy distribution (SED) curve slope is basically 
flat in the J and K-bands, but very sharp in the optical band.

\subsection{APT Photometry}
Photometry of HD~87646 was obtained in the Stromgren $b$ and $y$ bands 
between 2007 December and 2015 June with the T12 0.8-m Automatic 
Photoelectric Telescope (APT) at Fairborn Observatory in Arizona. 
Our primary goal with photometry is to detect if the companions transit  the primary 
star. The data were processed using software 
described in \citet{henry99}. The measurements have a typical 
accuracy of~$\sim$~0.001~mag.

\section{Results}

\subsection{Stellar Parameters}
\begin{figure}
\includegraphics[angle=0,width=8.2cm]{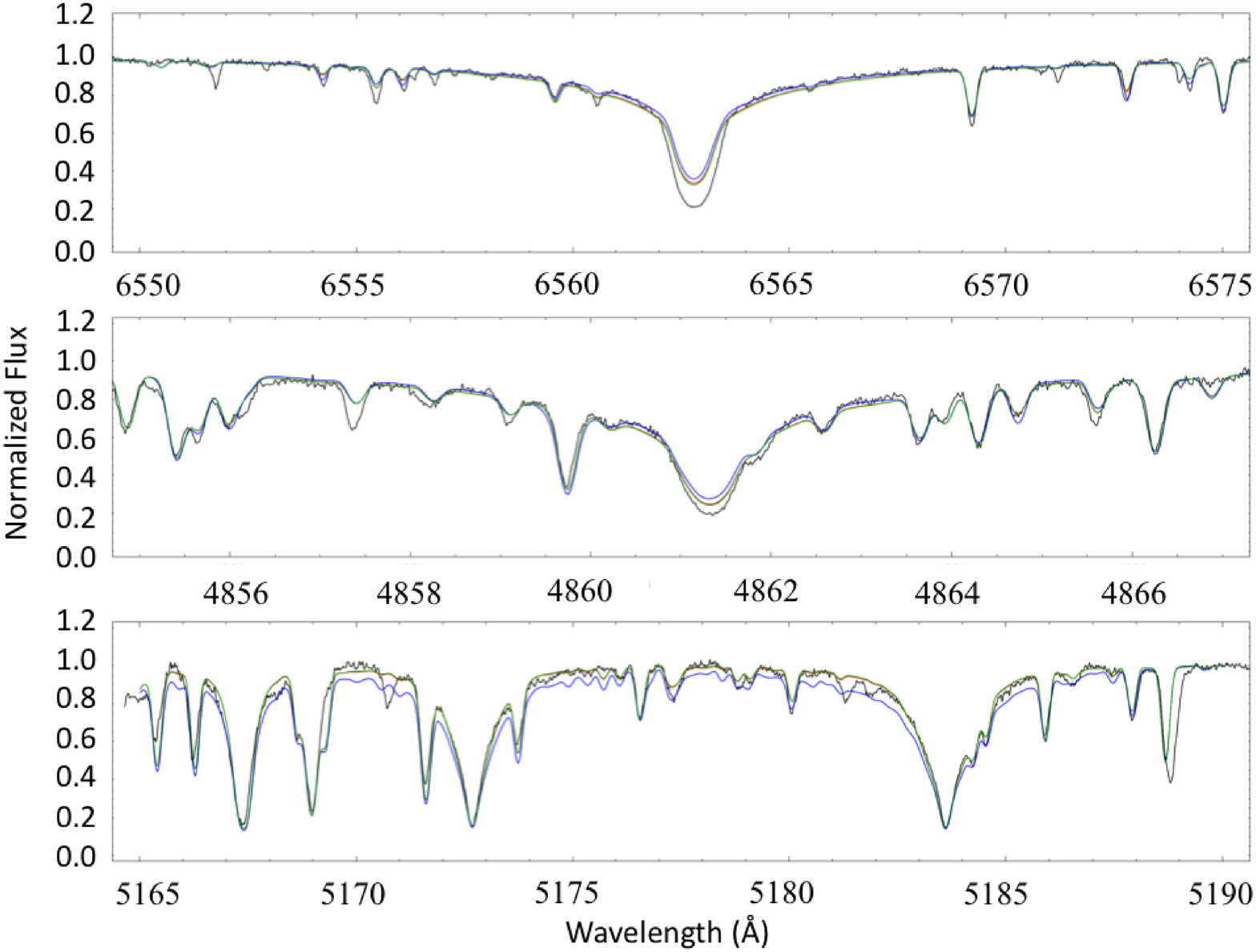}
\caption{High resolution TNG spectra of HD~87646 centered around H$\alpha$, H$\beta$ and Mgb lines (black lines).
The three other lines correspond to synthetic spectra for a binary with a G dwarf primary 
($T_{\rm eff}=5770$~K, \logg=4.1, [Fe/H] =-0.17, Microturbulence=1.8km~$\rm s^{-1}$ and $V\sin i$=7.5km~$\rm s^{-1}$) 
and a K dwarf secondary ($T_{\rm eff}=4000$~K, \logg=5.0). The green, red and blue lines 
correspond to a 0\%, 10\% and 50\% flux contribution from the secondary to the whole binary. }
\label{fig:tyc1415_synth}
\end{figure}

The SARG spectra taken at TNG without the iodine cell
were used to derive the stellar parameters. HD~87646 is flagged as a
binary in the Hipparcos and Tycho Catalogues \citep{perryman97}, with a Hipparcos magnitude (a broad band V filter)
difference between the primary and secondary to be 2.66$\pm$0.96~mag. 
Taking into account the binary nature of the object, 
we explored possible combinations of stellar parameters for the primary and
secondary. The SED and the colors change slightly due to the
secondary contribution; however, the normalized spectra show
minor changes, only affecting the wings of strong lines (e.g the Mgb lines, Fig.~\ref{fig:tyc1415_synth}).
The equivalent widths of most weak lines are essentially unchanged,
so we used Fe I and FeII lines with equivalent widths below 140m\AA\ and 
performed traditional spectroscopic analysis.

We use the latest MARCS model atmospheres
(Gustafsson et al. 2008) for the analysis. Generation
of synthetic spectra and the line analysis were performed
using the turbospectrum code (Alvarez \& Plez
1998), which employs line broadening according to the
prescription of Barklem \& O'Mara (1998). The line lists
used are drawn from a variety of sources. Atomic lines
are taken mainly from the VALD database
(Kupka et al. 1999). The molecular species CH, CN, OH,
CaH and TiO are provided by B. Plez (see Plez \& Cohen
2005), while the NH, MgH and C$_2$ molecules are from
the Kurucz line lists. The solar abundances used here
are the same as Asplund (2005). We use FeI excitation
equilibrium and derived an effective temperature
\teff = 5770$\pm$80K, which is slightly lower than  the effective
temperature derived from photometry (Feltzing \& Gustafsson 1998). 
The H$\alpha$ and  H$\beta$ wings also agree better for this lower \teff value.
We find log $\it{g}$=4.1$\pm$0.1, based on ionization equilibrium of FeI
and FeII lines and by fitting the wings of the Mgb lines, which agrees
with the previous estimates.  A microturbulence value of 1.8 km$\;$s$^{-1}$ is derived 
by forcing weak and strong FeI lines to yield the same abundances.
We are not able to confirm the super solar metallicity of this object 
(Feltzing \& Gustafsson 1998);  we derived [Fe/H] = $-0.17\pm0.08$.  
When we adopt the same \teff and microturbulence as 
Feltzing \& Gustafsson (1998), we obtain the same metallicity value [Fe/H]=0.3, 
but with a large slope in the excitation potential versus FeI abundance and 
reduced equivalent width versus FeI abundance. The derived stellar parameters are 
summarized in Table~\ref{tab:stellarparams}.

\begin{table}
\begin{center}
\caption{Parameters of the Star HD~87646A \label{tab:stellarparams}}
\begin{tabular}{lc}
\hline\hline
Parameter & Value \\
\hline
$T_{\rm eff}$ & 5770$\pm$80 K \\
$\log(g)$ & 4.1$\pm$0.1     \\
$\rm [Fe/H]$ &  -0.17$\pm$0.08    \\ 
 $V\sin i$ & 7.5 kms$^{-1}$ \\
 $\xi_{t}$ & 1.8 kms$^{-1}$ \\
Mass & 1.12 $\pm$ 0.09 $M_{\odot}$\\ 
Radius & $1.55\pm0.22 R_{\odot}$\\ 
\hline
\end{tabular}
\end{center}
\end{table}

We attempted to place constraints on the secondary star by fitting the 
Balmer line and Mgb line wings. Based on the Hipparcos data,
which suggest a Hipparcos magnitude difference between the primary and secondary
2.66$\pm$0.96~mag, the secondary has a flux contribution of 8-10\% 
with respect to the primary. We synthesized binary spectra with a G dwarf primary
and a K dwarf secondary, with 10\% and 50\% flux contribution from the secondary.
The binary spectrum synthesis results are consistent with the Hipparcos
data with 10\% contribution from the secondary. We cannot place better 
estimates based on the spectral line profiles because of the large errors
in the continuum normalization of these lines, since they spread over the 
entire echelle order. 

We use the empirical polynomial relations of \citet{Torres2010} to estimate 
the mass and radius of the primary star, HD~87646A, from $T_{\rm
eff}$, $\log(g)$, and [Fe/H]. These relations were derived from a sample of eclipsing
binaries with precisely measured masses and radii.  We estimate the
uncertainties in $M_*$ and $R_*$ by propagating the uncertainties in $T_{\rm
eff}$, $\log(g)$, and [Fe/H] using the covariance matrices of the 
\citet{Torres2010} relations (kindly provided by G. Torres). 
Since the polynomial relations of \citet{Torres2010} were derived empirically, 
the relations were subject to some intrinsic scatter, which we add in quadrature to the 
uncertainties propagated from the stellar parameter measurements 
\citep[$\sigma_{\log m} = 0.027$ and $\sigma_{\log r} = 0.014$;][]{Torres2010}. 
The final stellar mass and radius values obtained are 
$M_* = 1.12 \pm 0.09\,M_{\odot}$ and $R_* = 1.55\pm0.22\,R_{\odot}$.

\subsection{Systematic RV Errors Due to the Blended Binary Spectrum \label{sec:contamination}}
HD~87646 is a binary system, and contamination of the primary star's spectrum by 
the secondary star leads to an increased RV jitter by interfering with the analysis
pipeline. In this section we investigate the possible systematic
RV errors caused by the blended binary spectra using simulations. 
Since our RV observations were produced by two different kinds of spectrographs, 
we decided to perform two simulations, one for the DFDI instruments, including KeckET, 
KPNO ET, and MARVELS, and the other for traditional echelle 
spectrographs, including HRS at HET and the AST fiber-fed echelle 
spectrograph at Fairborn. In both simulations, we first create a set of stellar spectra 
by combining a G-type star (for the primary) and a K-type (for the secondary) star spectra with varying 
radial velocities for both stars. Then we calculate the differential radial 
velocities for the G-star from the simulated spectra. 
The differences between the output G-star RVs and input G-star RVs
are the RV errors caused by secondary star spectra contamination. 
Both simulations yield similar RV errors on the order of $200$~m~s$^{-1}$. 
We expect to see this level of systematic error and will include it in the RV `jitter' term 
when we perform the RV curve fitting in the next section. Traditionally the `jitter' term 
used to denote any RV noise caused by stellar activity; our `jitter' term also contains 
the RV noise caused by blended binary spectra.

\subsection{RV curve fitting and orbital parameters}
We have performed a Markov Chain Monte Carlo (MCMC) analyses of the 
combined RV observations from KeckET, ET, HET, MARVELS and Fairborn instruments. In this 
analyses, we initially used a one planet RV model to fit our RV observations, and 
later found that there is another strong periodic RV signal in the RV residuals. 
We then adopted a two object (a planet and a brown dwarf) RV model to fit our RV data. 
The RV model details are presented in \S 2 of 
\citet{gregory07}. We have attempted to add in another planet to fit our RV data 
around the third peak in the periodogram. The addition of another planet did not significantly improve our 
RV fitting. And combining with the fact that the newly added planet has a period half of the first 
giant planet, and eccentricity of 0.99, we consider it as an alias and over-fit of the noise. 
Thus we have rejected the RV model with a three substellar companions. 
Throughout the paper, we only present and discuss the RV model with two substellar companions.

Each state in the Markov chain is described by the parameter set
\begin{equation}
\vec{\theta}=\{P_1,K_1,e_1,\omega_1,M_1,P_2,K_2,e_2,\omega_2,M_2,
C_i,\sigma_{\rm jitter}\},
\end{equation}
where $P_1$ and $P_2$ are
orbital periods, $K_1$ and $K_2$ are the radial velocity semi-amplitudes, $e_1$ and $e_2$ 
are the orbital eccentricities, $\omega_1$ and $\omega_2$ are the arguments of periastron, 
$M_1$ and $M_2$ are the mean anomalies at chosen  epoch ($\tau$), $C_i$ is 
constant velocity offset between the differential RV data 
shown in Tables~\ref{tbl:KeckETvels}, \ref{tbl:KPNOvels}, \ref{tbl:HETvels}, \ref{tbl:MARVELS}, and \ref{tbl:Fairbornvels} 
and the zero-point of the Keplerian RV model 
($i=1$ for KeckET observations, $i=2$ for KPNO ET observations, $i=3$ for
HET observations, $i=4$ for MARVELS observations, and $i=5$ for Fairborn observations), 
and $\sigma_{\rm jitter}$ is the ``jitter'' parameter. 
The jitter parameter describes any excess noise, including both astrophysical
noise (e.g. stellar oscillation, stellar spots; Wright 2005),
any instrument noise not accounted for in the quoted measurement
uncertainties and systematic RV errors from analyzing blended binary spectra discussed in the last section. 
We use standard priors for each parameter 
(see Gregory 2007).  The prior is uniform in the logarithm of the orbital 
period ($P_1$ and $P_2$) from 1 to 5000 days.  
For $K_1$, $K_2$ and $\sigma_{\rm jitter}$ we use a modified Jefferys 
prior which takes the form of $p(x)=(x+x_o)^{-1}[\log(1+x_{max}/x_o]^{-1}$ 
, where $x_o=0.1$~${\rm m \, s^{-1}}$ and $x_{\mathrm
max} = 2128$~${\rm m \, s^{-1}}$ \citep{gregory05}.  
Priors for $e_1$ and $e_2$ are uniform between zero and unity. 
Priors for $\omega_1$, $\omega_2$, $M_1$ and $M_2$ are uniform 
between zero and $2\pi$.
For $C_i$, the priors are uniform between min($v_i$)-5\kms and  max($v_i$)+5\kms, 
where $v_i$ are the set of radial velocities obtained from each of the four instruments 
($i=1$ for KeckET observations, $i=2$ for KPNO ET observations, $i=3$ for
HET observations, $i=4$ for MARVELS observation, and $i=5$ for Fairborn observations). 
We verified that the chains did not approach the limiting value of $P_1$, 
$P_2$, $K_1$, $K_2$ and $\sigma_{\rm jitter}$. 

Following \citet{ford06}, we adopt a likelihood (i.e., conditional probability of making the specified 
measurements given a particular set of model parameters) of
\begin{equation}
p(v|\vec{\theta},M) \propto \prod_k \frac{\exp[-(v_{k,\theta}-v_k)^2/2(\sigma^2_{k,obs}+\sigma^2_{\rm jitter})]}{\sqrt{{\sigma_{k,obs}}^2+{\sigma_{\rm jitter}}^2}},
\end{equation}
where $v_k$ is observed radial velocity at time $t_k$, $v_{k,\theta}$ is the
model velocity at time $t_k$ given the model parameters
$\vec{\theta}$, and $\sigma_{k,obs}$ is the measurement uncertainty
for the radial velocity observation at time $t_k$.

We combine the Markov chains described above to estimate the joint
posterior probability distribution for the orbital model for
HD~87646. In Table~\ref{tbl:companionparams} we report the median
value and an uncertainty estimate for each model parameter based on
the marginal posterior probability distributions. The uncertainties
are calculated as the standard deviation about the mean value from
the combined posterior sample. Since the shape of the marginal
posterior distribution is roughly similar to a multivariate normal
distribution, the median value plus or minus the reported
uncertainty roughly corresponds to a 68.3\% credible interval. 
In the same table, we also reported the rms of the RV fitting residuals for the five different 
RV instruments used here, which are $\rm rms_1=245$~m~s$^{-1}$ for KeckET, 
$\rm rms_2=248$~m~s$^{-1}$ for KPNO ET, $\rm rms_3=261$~m~s$^{-1}$ for HET, 
$\rm rms_4=270$~m~s$^{-1}$ for MARVELS, 
and $\rm rms_5 = 312$~m~s$^{-1}$ for Fairborn RV observations.

\begin{figure}
\includegraphics[angle=0,width=8.2cm]{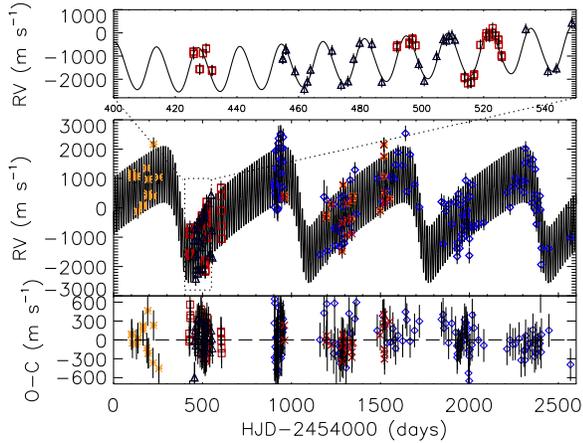}
\caption{{\it Top:} Expanded section of the middle panel plot 
(the dotted rectangular region) to show the short-period giant planet 
signal in the two-Keplerian RV model. 
{\it Middle:} Radial velocity observations of HD~87646 with the 
two-Keplerian model. 
{\it Bottom:} RV residuals of the two-Keplerian orbit model. 
Each panel shows radial velocity observations from KeckET (yellow stars), 
HET (black triangles), KPNO ET (red squares), Fairborn (blue diamonds), and 
MARVELS (red cross).} 
\label{fig:rv}
\end{figure}

\begin{figure}
\includegraphics[angle=0,width=8.2cm]{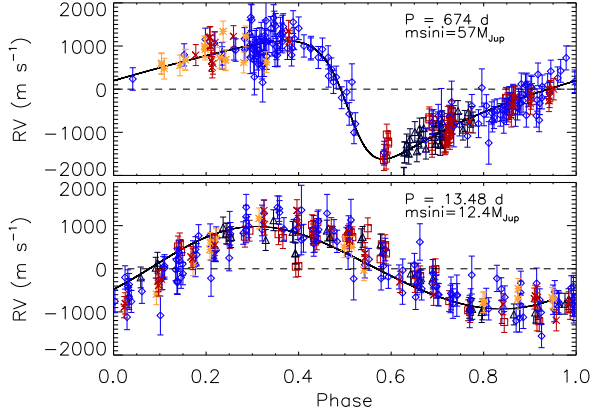}
\caption{Phased RV curves for the two signals in the two-Keplerian RV 
model. In each case, the contribution of the other signal was subtracted. 
Each panel shows radial velocity observations from KeckET (yellow stars), 
HET (black triangles), KPNO ET (red squares), Fairborn (blue diamonds), 
and MARVELS (red cross).} 
\label{fig:rvphased}
\end{figure}

\begin{table*}
\caption{{ Orbital Parameters for HD~87646b and HD~87646c  \label{tbl:companionparams}}}
\begin{tabular}{ccc}
\hline\hline
Parameter & HD~87646b & HD~87646c \\
\hline
Minimum Mass & 12.4$\pm$ 0.7 $M_{\rm Jup}$ & 57.0$\pm$ 3.7 $M_{\rm Jup}$\\
$a$ & 0.117 $\pm$ 0.003 AU & 1.58 $\pm$ 0.04 AU \\
$K$ & 956 $\pm$ 25 m~s$^{-1}$ & 1370$\pm$ 54 m~s$^{-1}$ \\
$P$ & 13.481 $\pm$ 0.001 d & 674 $\pm$ 4 d  \\
$e$ & 0.05$\pm$  0.02 & 0.50$\pm$  0.02 \\
$\omega$ (radians) & 5.20$\pm$ 0.52 & 1.95$\pm$0.06 \\
$T_{\rm{prediction\; for\; transit}}$ (JD$_{\rm UTC}$) & 2454093.85 $\pm$ 0.12 d &  \\ 
$T_{\rm periastron}$ (JD$_{\rm UTC}$) &  2454088.3$\pm1.1$ d & 2453707$\pm9$ d  \\
\hline
$\sigma_{\rm jitter}$ & 240 $\pm$ 12 m~s$^{-1}$  &  \\
$C_1$ & 20.878 $\pm$ 0.050 km~s$^{-1}$  &  \\
$C_2$ & 20.777 $\pm$ 0.090 km~s$^{-1}$  &  \\
$C_3$ & 0.908 $\pm$ 0.080 km~s$^{-1}$  &  \\
$C_4$ & -0.306 $\pm$ 0.050 km~s$^{-1}$  & \\
$C_5$ & 20.786 $\pm$ 0.070 km~s$^{-1}$  &  \\
$\rm rms_1$ & 245 m~s$^{-1}$  & \\
$\rm rms_2$ & 248 m~s$^{-1}$  &  \\
$\rm rms_3$ & 261 m~s$^{-1}$  & \\
$\rm rms_4$ & 270~m~s$^{-1}$  &   \\
$\rm rms_5$ & 312 m~s$^{-1}$  &  \\
\hline
\end{tabular}
\end{table*}
 
\begin{table*}
\caption{{ Orbital Parameters for HD~87646b and HD~87646c with a Linear RV Trend \label{tbl:companionparams2}}}
\begin{tabular}{ccc}
\hline\hline
Parameter & HD~87646b & HD~87646c \\
\hline
Minimum Mass & 12.4$\pm$ 0.7 $M_{\rm Jup}$ & 57.0$\pm$ 3.7 $M_{\rm Jup}$\\
$a$ & 0.117 $\pm$ 0.003 AU & 1.58 $\pm$ 0.04 AU \\
$K$ & 954 $\pm$ 24 m~s$^{-1}$ & 1370$\pm$ 56 m~s$^{-1}$ \\
$P$ & 13.481 $\pm$ 0.001 d & 673 $\pm$ 4 d  \\
$e$ & 0.05$\pm$  0.02 & 0.50$\pm$  0.02 \\
$\omega$ (radians) & 5.18$\pm$ 0.48 & 1.95$\pm$0.06 \\
$T_{\rm{prediction\; for\; transit}}$ (JD$_{\rm UTC}$) & 2454093.86 $\pm$ 0.14 d &  \\ 
$T_{\rm periastron}$ (JD$_{\rm UTC}$) &  2454088.2$\pm1.0$ d & 2453709$\pm8$ d  \\
\hline
$\rm{v_{trend}}$ &  $-12\pm18$  m~s$^{-1}$~yr$^{-1}$ & \\
$\sigma_{\rm jitter}$ & 240 $\pm$ 13 m~s$^{-1}$  &  \\
$C_1$ & 20.89 $\pm$ 0.070 km~s$^{-1}$  & \\
$C_2$ & 20.79 $\pm$ 0.10 km~s$^{-1}$  &  \\
$C_3$ & 0.92 $\pm$ 0.09 km~s$^{-1}$  &  \\
$C_4$ & -0.262 $\pm$ 0.06 km~s$^{-1}$  &  \\
$C_5$ & 20.838$\pm$ 0.09 km~s$^{-1}$  & \\
$\rm rms_1$ & 246 m~s$^{-1}$  &  \\
$\rm rms_2$ & 248 m~s$^{-1}$  & \\
$\rm rms_3$ & 261 m~s$^{-1}$  & \\
$\rm rms_4$ & 269~m~s$^{-1}$  &  \\
$\rm rms_5$ & 312 m~s$^{-1}$  &  \\
\hline
\end{tabular}
\end{table*}

This two-Keplerian orbital solution is shown in Figs~\ref{fig:rv} 
and \ref{fig:rvphased} together with the KeckET, HET, KPNO ET, Fairborn and MARVELS RV data. 
The residuals in these two plots can 
not be explained only by the errors in our RV data. A stellar jitter term 
$\sigma_{\rm jitter}=240\pm12$m~s$^{-1}$ is required in our fitting to 
explain these residuals. As discussed in the last section, most of 
the `jitter' noise arises from our data pipeline handling 
the blended binary spectra instead of a single star spectra. We also did an MCMC 
analysis using two `jitter' noise terms, one for DFDI instruments and the other one for 
echelle spectrographs, and find the orbital parameters for the giant planet candidate
and the BD candidate are barely changed within the error bars. 
So to keep it simple, we choose to use one `jitter' term for 
all our RV observations from different instruments.

HD~87646 is a binary system, so we have done another MCMC analysis by including a linear 
RV trend ($\rm{v_{trend}}\times(t-t_0)$) to the two-objects RV model used above. This linear RV 
trend is used to account for the perturbation of the primary star induced by the gravitational force 
of the secondary star. We note here that the offsets between different data sets will hinder the modeling 
of this linear trend as there is expected strong correlation between the offsets and this linear trend. 
Our new RV fitting yields orbital parameters for the two sub-stellar companions in 
addition to a linear RV trend of $\rm{v_{trend}}=-12\pm18$~m~s$^{-1}$~yr, which are 
summarized in Table~\ref{tbl:companionparams2}. Since all the main orbital parameters of 
the two sub-stellar companions are barely changed within their respect error bars and 
the strong correlation between this RV trend and telescopes RV offsets, we decide to 
keep using the numbers present in Table~\ref{tbl:companionparams} throughout this paper. 
This linear trend is not significant, which means it is more likely that either the secondary 
star is close to its ascending or descending node during 2008-2013, 
or this binary is on a relatively low-inclination (face on) orbit. It is not possible for us 
to distinguish these two scenarios using our current data. Future high precision astrometry 
observations, like GAIA, will help to solve this binary orbital problem. We also want 
to note here that this linear trend is not exact the real RV trend of the primary star induced by the gravitational
perturbation of the secondary star because of the secondary star's spectral contamination. It is close to
$\sim70$-$80\%$ of the real trend value assuming the flux ratio of the two stars is $\sim10$ in 
the optical band and mass ratio is $\sim2$.

\subsection{Line Bisector Analysis}
\citet{santos02} found small radial velocity variations and line asymmetries 
for the star HD~41004, which is a visual binary and is unresolved at 
the spectrograph. It was initially thought to have a planetary 
companion around the primary star, but from the line bisector analysis 
they were able to infer a possible brown dwarf orbiting the secondary star 
instead of a planet orbiting the primary star. Their conclusions were subsequently 
corroborated by \citet{zucker03}.

\begin{figure}
\includegraphics[angle=0,width=8.2cm]{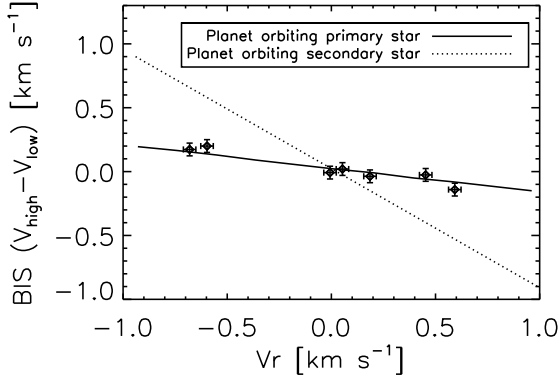}
\caption{Measured radial-velocity vs. BIS from EXPERT spectroscopic data for HD~87646. 
The solid and dotted lines show simulation results when assuming a 
giant planet orbiting the primary star and the secondary 
star, respectively. } 
\label{fig:bis}
\end{figure}

Similar to HD~41004, HD~87646 is also a binary system with a small angular 
separation ($0.4''$), which renders the spectrum a blended spectrum of the two 
stellar components. Following the same philosophy of 
\citet{santos02}, we performed a bisector analysis for HD~87646 to 
determine from which star in the binary system the RV signal was produced. 
We have analyzed spectra taken at the Kitt peak 2m telescope using 
EXPERT \citep{ge10}. Spectra were reduced using an IDL pipeline modified 
from an early version described in Wang (2012). Frames were trimmed, 
bias subtracted, flat-field corrected, aperture-traced, and extracted. 
Cross-Correlation Functions (CCFs) are derived by cross-correlation with a 
spectral mask from the wavelength range $4900-6300$\AA. Then we compute 
the bisector velocity for 10 different levels of the CCF. The values for the upper 
(near continuum) and lower bisector points are averaged and subtracted. 
The resulting quantity (the Bisector Inverse Slope, BIS) can be used to 
measure the line bisector variations \citep{queloz01}. 
The result of the bisector analysis is presented in Fig.~\ref{fig:bis}, 
which demonstrates that the BIS varies in phase with the radial velocity. 

We then created two simulations following \citet{santos02}, one by assuming 
a giant planet orbiting the primary star of the binary (solid line in Fig.~\ref{fig:bis}) 
and the other by assuming a heavier BD orbiting the secondary star (dotted 
line in Fig.~\ref{fig:bis}). Both scenarios can explain the RV curve seen 
for HD~87646, but clearly only the one in which the giant planet is orbiting 
the primary star (solid line in Fig.~\ref{fig:bis}) is consistent with the BIS analysis. 
Our conclusion from the BIS analysis is that the 13.5-day period giant planet is 
orbiting the primary star HD~87646A. 

\subsection{Photometry results}

\begin{figure}
\includegraphics[angle=0,width=8.2cm]{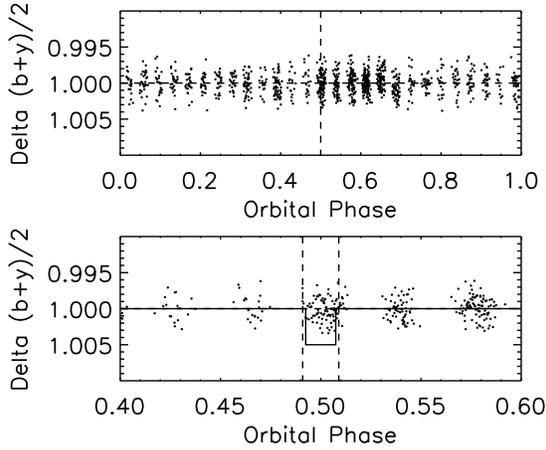}
\caption{Top: the 1077 differential magnitudes of HD~87646 phased to the period 
of the giant planet HD~87646Ab, taken by the 0.8m APT from 2008-2015. 
The horizontal dashed line corresponds to the mean brightness level of the
1077 observations. The vertical dashed line marks the expected time of mid-transit.
Bottom: an expanded portion of the top plot, centered on the predicted central transit window. 
The solid curve shows the predicted central transit, with a depth of 0.005 mags 
and duration of 0.015 units of phase. The $\pm1\sigma$ uncertainty in the transit window timing is 
indicated by the two vertical dotted lines.
} 
\label{fig:photometry}
\end{figure}

The top panel of Fig.~\ref{fig:photometry} presents all 1077 photometric observations plotted 
against the latest transit ephemeris of HD~87646a:  $T_{\rm c} = 2454093.85$~d, $ P= 13.481$~d.  
The differential magnitudes are measured against the mean of three comparison stars 
to improve precision. The standard deviation of these data from their mean 
is 0.0014 mag. A least-squares sine-curve fit to the phased data yields a full amplitude of 
0.000089 $\pm$0.000065 mag. There is no detectable brightness variation 
on the short period radial velocity period; this result supports the interpretation that the observed 
radial velocity variations are caused by a companion.

The bottom panel of Fig.~\ref{fig:photometry} is similar to the top panel except that 
it displays only the data within $\pm 0.1$ phase units from the predicted transit time. 
We also show the predicted central transit, phased at 0.5, for a duration of 0.21 days 
or $\sim0.015$ units of phase and a depth of $0.5\%$ or $\sim0.005$~mag \citep{kane08}. 
The $\pm1\sigma$ uncertainty in the transit window timing is indicated by the two 
vertical dotted lines. There are 1005 observation that lie outside the predicted transit window, 
which have a mean of 0.99998$\pm$0.00005~mag and 72 observations that 
fell within the transit window have a mean of 1.0003$\pm$0.0002~mag.
The difference between these two mean brightness is $0.0003\pm0.0002$~mag.  
Full transits with a predicted depth $\sim0.005$~mag are excluded by the photometry 
at the predicted transit time.

\subsection{Companion Inclination \& Mass Estimate}
The mass function is related to the observed period,
eccentricity and radial velocity semi-amplitude as:
\begin{equation}
  \frac{(m \sin{i})^3}{(M_*+m)^2} = \frac{P(1-e^2)^{\frac{3}{2}}K^3}{2\pi G}
  \label{eqn:massfunction}
\end{equation}
where $M_*$ is the mass of the primary and $m$ the mass of the
companions. Since the first companion is known not to transit the star, we cannot break the 
degeneracy of mass and $\sin{i}$ with radial velocity observations alone. 
Using the derived stellar mass (1.12$M_\odot$) for the primary with the 
orbital parameters determined from the radial velocity (Table~\ref{tbl:companionparams}) 
we determine that the minimum mass of the inner companion for an edge-on orbit ($\sin{i} = 1$) 
is 12.4$\pm$0.7~M$_{\rm Jup}$. This mass is quite close to the deuterium burning 
limit, and the detected companion is likely burning deuterium, although its minimum mass places 
it in the giant planet regime. The second companion's minimum mass when assuming an edge-on orbit 
is 57.0$\pm3.7$ M$_{\rm Jup}$, which falls right into the brown dwarf regime.

\section{Summary and Discussion}
\subsection{Summary of the main results}

Our SDSS MARVELS pilot survey and additional observations at the HET, 
KPNO 2.1m telescope, and Fairborn observatory confirm the detection of two massive 
substellar companions in a close binary system HD~87646. 
The first companion, HD~87646Ab, has a minimum mass of 12.4$\pm$0.7 M$_{\rm Jup}$, 
period of 13.481$\pm$0.001 days and eccentricity of 0.05$\pm$0.02. The measured 
eccentricity is in line with other short period giant planets in binaries 
\citep[e.g.,][]{Eggenberger07}. This companion is likely to be a giant planet or 
a brown dwarf, depending on its inclination angle. 
Our bisector analysis has shown that this companion is in orbit around the primary star. 
The second companion has a minimum mass of 57$\pm3.7$ M$_{\rm Jup}$, period of 
674$\pm$4 days and eccentricity of 0.50$\pm$0.02. This companion is likely 
to be a brown dwarf. \citet{ma14} have found long period high mass 
($>$42M$_{\rm Jup}$) BDs tend to have higher eccentricities. This new BD 
is consistent with this trend.  

This is the eleventh detection of a substellar companion(s) in a binary system with 
separation of only about 20~AU. The other ten systems are Gliese~86 
(Queloz et al. 2000; Mugrauer \& Newhauser 2005; Lagrange et al. 2006), 
$\gamma$ Cephei (Hatzes et al. 2003), HD~41004 (Zucker et al. 2003, 2004), 
HD~188753 \citep{Konacki05}, HD~176051 \citep{muterspaugh10}, HD~126614 \citep{ho10}, 
$\alpha$ Centauri \citep{dumusque12}, HD~196885\citep{correia08}, OGLE-2013-BLG-0341 \citep{gould14} 
and HD~59686 \citep{ortiz16}. 
However, \citet{Eggenberger07} did not confirm the planet in HD~188753, and 
\citet{raj16} suggest the planet signal discovered from $\alpha$ Centauri B is 
not from a real planet, but from the observation window function. To the best of our knowledge, 
HD~87646A is the first multiple planet/BD system detected in such close binaries. 

\subsection{Dynamical Stability of HD~87646}

\begin{figure}
\includegraphics[angle=0,width=8.2cm]{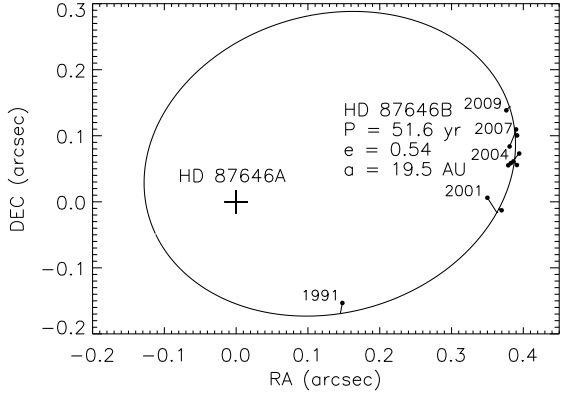}
\caption{Orbital data for binary HD~87646. The plus symbol marks the location of 
the primary (HD~87646A), filled circles are measured position of HD~87646B 
from literature and this paper, and line segments are drawn from the ephemeris prediction 
to the observed location of the secondary in each case. } 
\label{fig:binary_sepa}
\end{figure}

In this section we will discuss the dynamical stability of the binary system HD~87646. 
First we have collected observational data of HD~87646 from the literature (Horch et al. 2008, 
Hartkopf \& Mason 2009, Horch et al. 2010, Balega et al. 2013) and combined them with our 
AO data to constrain the binary orbit of HD~87646. Our best fitting binary orbital 
solution (solid line) and the observational data (black dots) are shown in 
Fig.~\ref{fig:binary_sepa}. The best fitting parameters are $P = 51.6$~yr, 
$e=0.54$ and $a=0.26$~arcsec. At a distance of $73.58\pm9.68$~pc, 
this angular separation corresponds to a binary semi-major axis of $19\pm2$~AU. 
Since the observational data only cover less than half of the binary orbit, these 
fitting parameters are very preliminary. There are many previous cases 
for which binary orbital parameters were revised significantly with 
new and better astrometry measurements, especially when the binary orbit is 
not completely covered by astrometry observations yet \citep{hartkopf09}. 
Using the same classification of visual binaries as that in \citet{hartkopf01} Hartkopf et al. (2001) for the 
Catalog of Visual Binary, our orbital solution has a grade 4 ( 1 = definitive, 2 = good, 
3 = reliable, 4 = preliminary, 5 = indeterminate) and formal errors of the orbital 
solution were considered to have little meaning. 

\begin{figure}
\includegraphics[angle=0,width=8.2cm]{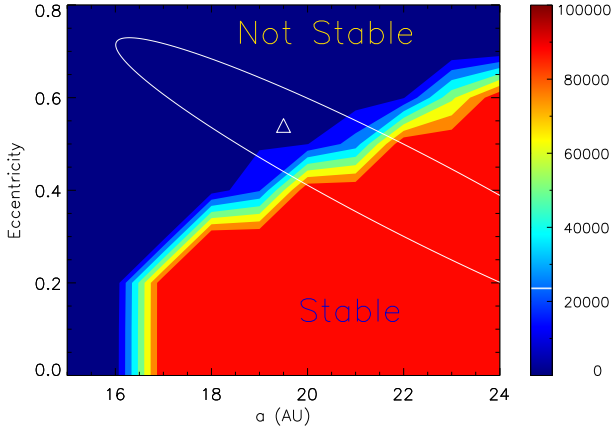}
\caption{Dynamical simulation results for HD~87646. The contour lines show the time the 
system will remain stable according to our Mecury simulation. The unit of the color bar is years. 
A stable zone is found in the eccentricity-semimajor axis diagram, which 
shows that if the binary orbit has a large semi-major axis and low eccentricity, 
the system will remain stable. The white triangle symbol shows the best binary orbital fit 
from the current astrometry data with $\chi^{2}_{\rm red}=1$ after we rescale the error bars for 
astrometric data (see also Fig.~\ref{fig:binary_sepa}). The white ellipse shows shows the distribution of binary orbital 
parameters from fitting the astrometric data fitting with max $\chi^{2}_{\rm red}=2$. There is an overlap region between the 
distribution of binary orbital parameters and the dynamically stable region, which demonstrates 
the binary system has stable orbital solutions.} 
\label{fig:dynamic}
\end{figure}

We then performed a numerical simulation of the binary system including the 
giant planet and brown dwarf discovered in this paper. The binary-planetary-brown dwarf system 
of HD~87646 was integrated numerically using the Bulirsch-Stoer integrator of 
the N-body integration package Mercury \citep{chambers99}. For each simulation, we tested the 
dynamical stability of this system given the semi-major axis (a$_{\rm B}$) and 
eccentricity (e$_{\rm B}$) of the binary system up to a million years. 
We have assumed the giant planet, brown dwarf and binary to be coplanar. 
The results are shown in Fig.~\ref{fig:dynamic}. 
There is a stable zone in the binary a$_{\rm B}$-e$_{\rm B}$ diagram. 
We have scaled the error bars of these astrometry data to force the best orbital fit to have a reduced 
chi-squared $\chi^{2}_{\rm red}=1$ and then over-plotted the binary orbital parameter fit from astrometry data 
with max $\chi^{2}_{\rm red}=2$ in Fig.~\ref{fig:dynamic}. The big uncertainty of the binary orbital 
fitting from astrometry data arises from the big error bar ($\sim0.1$~arcsec) of the 1991 Hipparcos data point. 
We want to point out there is another caveat of this plot, which is from our 
scaling of the error bars of all the astrometry data points. These data are from 5 different previous 
observation programs and scaling all of them together to make the best fit 
have a $\chi^{2}_{\rm red}=1$ can be problematic. 
The main conclusion from the simulation study and astrometry data fitting is, 
with a large binary semi-major axis (a$_{\rm B} > 17$ AU) and a relatively low binary 
eccentricity ($\rm e_{B} < (a_{\rm B} -17)\times0.57+0.2$), the binary-planet-brown 
dwarf system discovered in this paper is stable.

\subsection{Nature of the system and its formation and evolution}

HD~87646 is the first known system to have two massive substellar objects orbiting 
a star in a close binary system. Interestingly, the masses of these two substellar 
objects are close to the minimum masses for burning deuterium ($\sim$13 M$_{\rm Jup}$, 
Spiegel et al. 2011) and hydrogen ($\sim$75M$_{\rm Jup}$, Chabrier et al. 2000) which 
are generally assumed to be the general mass boundaries between planet and brown dwarf 
and between brown dwarf and star, respectively. All these peculiarities raise a 
question: how could such a system be formed? Here we briefly discuss this intriguing issue.

The large masses of these two substellar objects suggest that they could 
be formed as stars with their binary hosts: a large molecular cloud collapsed and fragmented 
into four pieces; the larger two successfully became stars and formed the HD~87646 binary, 
and the other smaller ones failed to form stars and became the substellar objects 
in this system \citep{chabrier14}. This scenario might be relevant for the binary 
stars but seems problematic for the two substellar objects on orbits within $\sim$1~AU because 
it is unclear whether fragmentation on such a small scale can occur \citep{kratter11}.

Perhaps a more plausible explanation is that the two substellar objects were formed like giant 
planet in a protoplanetary disk around HD~87646A. As for giant planet formation, there are currently 
two main models: core accretion vs. disk instability \citep[see the review by][]{angelo10, helled14}. 
Recently, many studies have examined the 
core accretion model's ability to form planets in close binaries with separations of $\sim$20 AU 
 \citep[see the review by][and the references therein]{thebault14}.  
A commonly recognized issue is that the binary perturbations generally inhibit the growth 
of planetesimals in the disk \citep{thebault11}. Even if their growth could proceed in certain 
favorable conditions \citep{xie08,xie09}, it would be significantly slowed, requiring 
a time scale of 10$^6$ yr or even longer (Xie et al. 2010).   This result raises a problem 
for the formation of a gaseous giant planet, as it would not form a planetary core 
(via planetesimal growth) to accrete gas before the gas disk dissipation, which takes a timescale 
as short as $10^5$-$10^6$ yr for such close binaries \citep{cieza09, kraus12}. 
Following the above logic, \citet{xie10} found that Jupiter-like planets are unlikely to 
form around Alpha Centauri B. As for the case of HD~87646, the formation of the two 
massive substellar objects via the core accretion model would be more problematic because 
it requires a more massive disk with mass larger than 68 M$_{\rm Jup}$. Such a massive disk is 
seldom observed in close binaries, which indicates that should such a massive disk exist, 
it would dissipate much faster than a normal lighter one. 

Conversely, the disk instability model could circumvent most of the above 
barriers. First, disk instability usually requires very high disk mass, which is in line with 
the masses of the two detected substellar objects. Second, planet formation via disk 
instability requires a short timescale, which is also consistent with the short disk dissipation 
timescale observed in close binaries. In addition, the model of disk instability is recently 
advocated by \citet{duchene10}, who argued that planet formation might be dominated by 
disk instability in close binaries based on the fact that exoplanets within close binaries 
(separation $<$ 100 AU) are significantly more massive than those within wide binaries or single stars. 
We could use the planet and brown dwarf mass to estimate the minimum surface
density of the primordial disk, and test if such a disk is gravitationally unstable.
We adopt the similarity solution of the evolving viscous disk \citep{hartmann98} where the surface 
density follows R$^{-1}$ from the star to the disk edge. Since the binary separation is 19~AU, 
the tidal truncation radius for the circumstellar disk around the primary star is $\sim6$~AU (1/3 of the binary separation).
Spreading the total mass of 12.4+57 Jupiter mass to this 6 AU disk, the disk surface 
density at 6~AU is 2604~g/cm$^2$. At 6 AU, the temperature is around 90 K 
with 1 solar luminosity. The sound speed is 0.56 km/s. Then the Toomre Q parameter with a 1.12 solar mass star is 1.5. 
At 1 AU, the temperature is around 220 K. The surface density is 15610~g/cm$^2$.
The sound speed is 0.87 km/s. The Toomre Q parameter is 5.7. Considering the
primordial disk mass should be a lot more massive than the planet and brown dwarf mass,
the disk is likely to be gravitationally unstable throughout the disk. This is consistent
with gravitational instability leading to planet formation.
Although several advantages exist for the disk instability model, we consider  that such an 
explanation for the formation of the HD~87646 system should be taken with caution because 
whether disk instability can be triggered in the present of a close stellar companion remains 
an issue under debate \citep{nelson00b, mayer05, boss06}.  

Next we want to discuss how the giant planet, b, moved to its current position with a very low eccentricity. 
Although the Kozai-Lidov mechanism and subsequent tidal dissipation \citep{wu03, fabrycky07} 
have been invoked often to explain the formation of hot Jupiters, it is unlikely that the brown dwarf, c, has helped to 
move b inward because such a process cannot help to form `warm' Jupiters \citep{antonini16}. 
So we prefer a scenario in which b initially formed in the disk and then migrated inward in the disk to its current position, 
which explains why it has a near zero eccentricity. As for the brown dwarf, c, 
after it formed in the disk, scattering between c and other objects formed in the disk moved it to a higher eccentricity. 
During such a scattering process, lower mass objects tend to be ejected out 
and more massive objects are kicked inward with a higher eccentricity according to the simulation of \citet{chatterjee08}. 
The stellar companion, B, cannot excite the eccentricity of c because  Kozai oscillations 
will be destroyed by the presence of other massive sub-stellar objects in the system (in this case, the giant planet b) 
according to the studies of \citet{wu03} and \citet{fabrycky07}. But the presence of the stellar companion, B, 
can help to enhance the scattering process between c and other objects formed in the protoplanetary 
disk around A \citep{marzari05}. Future astrometry observations, like those from Gaia \citep{Perryman01}, 
can provide a better binary orbital solution and even possibly constrain the BD candidate's orbit. 
These data will help to study the dynamic structure of this complicated system and give more 
insight to its formation scenario.

\acknowledgments

Funding for the multi-object Doppler instrument was provided by the  W.M. Keck Foundation. 
The pilot survey was funded by NSF with grant AST-0705139, NASA with grant NNX07AP14G 
and the University of Florida.
Funding for SDSS-III has been provided by the Alfred P. Sloan Foundation, the Participating Institutions, the National Science Foundation, and the U.S. Department of Energy Office of Science. The SDSS-III web site is http://www.sdss3.org/.
SDSS-III is managed by the Astrophysical Research Consortium for the Participating Institutions of the SDSS-III Collaboration including the University of Arizona, the Brazilian Participation Group, Brookhaven National Laboratory, Carnegie Mellon University, University of Florida, the French Participation Group, the German Participation Group, Harvard University, the Instituto de Astrofisica de Canarias, the Michigan State/Notre Dame/JINA Participation Group, Johns Hopkins University, Lawrence Berkeley National Laboratory, Max Planck Institute for Astrophysics, Max Planck Institute for Extraterrestrial Physics, New Mexico State University, New York University, Ohio State University, Pennsylvania State University, University of Portsmouth, Princeton University, the Spanish Participation Group, University of Tokyo, University of Utah, Vanderbilt University, University of Virginia, University of Washington, and Yale University.
The research at Tennessee State University was made possible 
by NSF support through grant 1039522 of the Major Research 
Instrumentation Program.  In addition, astronomy at Tennessee State 
University is supported by the state of Tennessee through its 
Centers of Excellence programs.

\end{CJK}

\begin{thebibliography}{}

\bibitem[Alam et al.(2015)]{alam15} Alam, S., Albareti, F.~D., 
Allende Prieto, C., et al.\ 2015, \apjs, 219, 12

\bibitem[{{Alvarez} {et~al}\mbox{.}(1998){Alvarez}, {Hernandez}, {Michel},
  {Jiang}, {Belmonte}, {Chevreton}, {Massacrier}, {Liu}, {Li}, {Goupil},
  {Cortes}, {Mangeney}, {Dolez}, {Valtier}, {Vidal}, {Sperl}, \&
  {Talon}}]{alvarez1998}
{Alvarez} M. {et~al.}, 1998, \aap, 340, 149

\bibitem[Antonini et al.(2016)]{antonini16} Antonini, F., Hamers, A.~S., \& Lithwick, Y.\ 2016, arXiv:1604.01781 

\bibitem[D'Angelo et al.(2010)]{angelo10} D'Angelo, G., Durisen, 
R.~H., \& Lissauer, J.~J.\ 2010, Exoplanets, 319

\bibitem[{{Asplund}(2005)}]{asplund05}
{Asplund} M., 2005, \araa, 43, 481

\bibitem[{{Balega} {et~al}\mbox{.}(2013){Balega}, {Balega}, {Gasanova},
  {Dyachenko}, {Maksimov}, {Malogolovets}, {Rastegaev}, \&
  {Shkhagosheva}}]{balega13}
{Balega} I.~I., {Balega} Y.~Y., {Gasanova} L.~T., {Dyachenko} V.~V., {Maksimov}
  A.~F., {Malogolovets} E.~V., {Rastegaev} D.~A., {Shkhagosheva} Z.~U., 2013,
  Astrophysical Bulletin, 68, 53

\bibitem[{{Barklem} \& {O'Mara}(1998)}]{barklem98}
{Barklem} P.~S., {O'Mara} B.~J., 1998, \mnras, 300, 863

\bibitem[{{Boss}(2006)}]{boss06}
{Boss} A.~P., 2006, \apj, 643, 501

\bibitem[{{Butler} {et~al}\mbox{.}(2004){Butler}, {Marcy}, {Fischer}, {Vogt},
  {Tinney}, {Jones}, {Penny}, \& {Apps}}]{Butler04}
{Butler} R.~P., {Marcy} G.~W., {Fischer} D.~A., {Vogt} S.~S., {Tinney} C.~G.,
  {Jones} H.~R.~A., {Penny} A.~J., {Apps} K., 2004, in IAU Symposium, Vol. 202,
  Planetary Systems in the Universe, {Penny} A., ed., p.~3

\bibitem[{{Chabrier} \& {Baraffe}(2000)}]{chabrier00}
{Chabrier} G., {Baraffe} I., 2000, \araa, 38, 337

\bibitem[{{Chabrier} {et~al}\mbox{.}(2014){Chabrier}, {Johansen}, {Janson}, \&
  {Rafikov}}]{chabrier14}
{Chabrier} G., {Johansen} A., {Janson} M., {Rafikov} R., 2014, Protostars and
  Planets VI, 619

\bibitem[Chambers(1999)]{chambers99} Chambers, J.~E.\ 1999, 
\mnras, 304, 793

\bibitem[Chatterjee et al.(2008)]{chatterjee08} Chatterjee, S., Ford, E.~B., Matsumura, S., \& Rasio, F.~A.\ 2008, \apj, 686, 580-602

\bibitem[{{Cieza} {et~al}\mbox{.}(2009){Cieza}, {Padgett}, {Allen}, {McCabe},
  {Brooke}, {Carey}, {Chapman}, {Fukagawa}, {Huard}, {Noriga-Crespo},
  {Peterson}, \& {Rebull}}]{cieza09}
{Cieza} L.~A. {et~al.}, 2009, \apjl, 696, L84

\bibitem[Correia et al.(2008)]{correia08} Correia, A.~C.~M., Udry, S., Mayor, M., et al.\ 2008, \aap, 479, 271 

\bibitem[{{Duch{\^e}ne}(2010)}]{duchene10}
{Duch{\^e}ne} G., 2010, \apjl, 709, L114

\bibitem[{{Dumusque} {et~al}\mbox{.}(2012){Dumusque}, {Pepe}, {Lovis},
  {S{\'e}gransan}, {Sahlmann}, {Benz}, {Bouchy}, {Mayor}, {Queloz}, {Santos},
  \& {Udry}}]{dumusque12}
{Dumusque} X. {et~al.}, 2012, \nat, 491, 207

\bibitem[{{Eaton} \& {Williamson}(2004)}]{eaton04}
{Eaton} J.~A., {Williamson} M.~H., 2004, in Society of Photo-Optical
  Instrumentation Engineers (SPIE) Conference Series, Vol. 5496, Advanced
  Software, Control, and Communication Systems for Astronomy, {Lewis} H.,
  {Raffi} G., eds., pp. 710--717

\bibitem[{{Eaton} \& {Williamson}(2007)}]{eaton07}
{Eaton} J.~A., {Williamson} M.~H., 2007, \pasp, 119, 886

\bibitem[{{Eggenberger} {et~al}\mbox{.}(2007){Eggenberger}, {Udry}, {Mazeh},
  {Segal}, \& {Mayor}}]{Eggenberger07}
{Eggenberger} A., {Udry} S., {Mazeh} T., {Segal} Y., {Mayor} M., 2007, \aap,
  466, 1179

\bibitem[{{Eisenstein} {et~al}\mbox{.}(2011){Eisenstein}, {Weinberg}, {Agol},
  {Aihara}, {Allende Prieto}, {Anderson}, {Arns}, {Aubourg}, {Bailey},
  {Balbinot}, \& et~al.}]{eisenstein11}
{Eisenstein} D.~J. {et~al.}, 2011, \aj, 142, 72

\bibitem[{{Erskine}(2003)}]{Erskine03b}
{Erskine} D.~J., 2003, \pasp, 115, 255

\bibitem[{{Erskine} \& {Ge}(2000)}]{erskine00}
{Erskine} D.~J., {Ge} J., 2000, in Astronomical Society of the Pacific
  Conference Series, Vol. 195, Imaging the Universe in Three Dimensions, {van
  Breugel} W., {Bland-Hawthorn} J., eds., p. 501

\bibitem[Fabrycky \& Tremaine(2007)]{fabrycky07} Fabrycky, D., \& Tremaine, S.\ 2007, \apj, 669, 1298 

\bibitem[{{Fekel} {et~al}\mbox{.}(2013){Fekel}, {Rajabi}, {Muterspaugh}, \&
  {Williamson}}]{fekel13}
{Fekel} F.~C., {Rajabi} S., {Muterspaugh} M.~W., {Williamson} M.~H., 2013, \aj,
  145, 111

\bibitem[{{Feltzing} \& {Gustafsson}(1998)}]{feltzing98}
{Feltzing} S., {Gustafsson} B., 1998, \aaps, 129, 237

\bibitem[{{Fischer} \& {Valenti}(2005)}]{Fischer05}
{Fischer} D.~A., {Valenti} J., 2005, \apj, 622, 1102

\bibitem[{{Fleming} {et~al}\mbox{.}(2010){Fleming}, {Ge}, {Mahadevan}, {Lee},
  {Eastman}, {Siverd}, {Gaudi}, {Niedzielski}, {Sivarani}, {Stassun},
  {Wolszczan}, {Barnes}, {Gary}, {Nguyen}, {Morehead}, {Wan}, {Zhao}, {Liu},
  {Guo}, {Kane}, {van Eyken}, {De Lee}, {Crepp}, {Shelden}, {Laws},
  {Wisniewski}, {Schneider}, {Pepper}, {Snedden}, {Pan}, {Bizyaev},
  {Brewington}, {Malanushenko}, {Malanushenko}, {Oravetz}, {Simmons}, \&
  {Watters}}]{fleming10}
{Fleming} S.~W. {et~al.}, 2010, \apj, 718, 1186

\bibitem[{{Ford}(2006)}]{ford06}
{Ford} E.~B., 2006, \apj, 642, 505

\bibitem[Perryman et al.(2001)]{Perryman01} Perryman, M.~A.~C., de Boer, 
K.~S., Gilmore, G., et al.\ 2001, \aap, 369, 339 

\bibitem[{{Ge}(2002)}]{ge02}
{Ge} J., 2002, \apjl, 571, L165

\bibitem[{{Ge}, {Erskine} \& {Rushford}(2002){Ge}, {Erskine}, \&
  {Rushford}}]{ge02b}
{Ge} J., {Erskine} D.~J., {Rushford} M., 2002, \pasp, 114, 1016

\bibitem[{{Ge} {et~al}\mbox{.}(2003){Ge}, {Mahadevan}, {van Eyken}, {Dewitt},
  \& {Shaklan}}]{ge03}
{Ge} J., {Mahadevan} S., {van Eyken} J., {Dewitt} C., {Shaklan} S., 2003, in
  Astronomical Society of the Pacific Conference Series, Vol. 294, Scientific
  Frontiers in Research on Extrasolar Planets, {Deming} D., {Seager} S., eds.,
  pp. 573--580

\bibitem[{{Ge} {et~al}\mbox{.}(2006{\natexlab{a}}){Ge}, {van Eyken},
  {Mahadevan}, {DeWitt}, {Kane}, {Cohen}, {Vanden Heuvel}, {Fleming}, {Guo},
  {Henry}, {Schneider}, {Ramsey}, {Wittenmyer}, {Endl}, {Cochran}, {Ford},
  {Mart{\'{\i}}n}, {Israelian}, {Valenti}, \& {Montes}}]{ge06}
{Ge} J. {et~al.}, 2006{\natexlab{a}}, \apj, 648, 683

\bibitem[{{Ge} {et~al}\mbox{.}(2006{\natexlab{b}}){Ge}, {Wan}, {Zhao},
  {Hariharan}, {Mahadevan}, {van Eyken}, {Guo}, {McDavitt}, {DeWitt}, {Cohen},
  {Fleming}, {Kane}, {Crepp}, \& {Shaklan}}]{ge06c}
{Ge} J. {et~al.}, 2006{\natexlab{b}}, in Society of Photo-Optical
  Instrumentation Engineers (SPIE) Conference Series, Vol. 6269, Society of
  Photo-Optical Instrumentation Engineers (SPIE) Conference Series, p.~2

\bibitem[{{Ge} {et~al}\mbox{.}(2010){Ge}, {Zhao}, {Groot}, {Chang}, {Varosi},
  {Wan}, {Powell}, {Jiang}, {Hanna}, {Wang}, {Pais}, {Liu}, {Dou}, {Schofield},
  {McDowell}, {Costello}, {Delgado-Navarro}, {Fleming}, {Lee}, {Bollampally},
  {Bosman}, {Jakeman}, {Fletcher}, \& {Marquez}}]{ge10}
{Ge} J. {et~al.}, 2010, in Society of Photo-Optical Instrumentation Engineers
  (SPIE) Conference Series, Vol. 7735, Society of Photo-Optical Instrumentation
  Engineers (SPIE) Conference Series, p.~0

\bibitem[{{Gonzalez} {et~al}\mbox{.}(2001){Gonzalez}, {Zaritsky}, {Dalcanton},
  \& {Nelson}}]{Gonzalez01}
{Gonzalez} A.~H., {Zaritsky} D., {Dalcanton} J.~J., {Nelson} A., 2001, \apjs,
  137, 117

\bibitem[{{Gould} {et~al}\mbox{.}(2014){Gould}, {Udalski}, {Shin}, {Porritt},
  {Skowron}, {Han}, {Yee}, {Koz{\l}owski}, {Choi}, {Poleski}, {Wyrzykowski},
  {Ulaczyk}, {Pietrukowicz}, {Mr{\'o}z}, {Szyma{\'n}ski}, {Kubiak},
  {Soszy{\'n}ski}, {Pietrzy{\'n}ski}, {Gaudi}, {Christie}, {Drummond},
  {McCormick}, {Natusch}, {Ngan}, {Tan}, {Albrow}, {DePoy}, {Hwang}, {Jung},
  {Lee}, {Park}, {Pogge}, {Abe}, {Bennett}, {Bond}, {Botzler}, {Freeman},
  {Fukui}, {Fukunaga}, {Itow}, {Koshimoto}, {Larsen}, {Ling}, {Masuda},
  {Matsubara}, {Muraki}, {Namba}, {Ohnishi}, {Philpott}, {Rattenbury}, {Saito},
  {Sullivan}, {Sumi}, {Suzuki}, {Tristram}, {Tsurumi}, {Wada}, {Yamai}, {Yock},
  {Yonehara}, {Shvartzvald}, {Maoz}, {Kaspi}, \& {Friedmann}}]{gould14}
{Gould} A. {et~al.}, 2014, Science, 345, 46

\bibitem[{{Gregory}(2005)}]{gregory05}
{Gregory} P.~C., 2005, \apj, 631, 1198

\bibitem[{{Gregory}(2007)}]{gregory07}
{Gregory} P.~C., 2007, \mnras, 381, 1607

\bibitem[{{Gunn} {et~al}\mbox{.}(2006){Gunn}, {Siegmund}, {Mannery}, {Owen},
  {Hull}, {Leger}, {Carey}, {Knapp}, {York}, {Boroski}, {Kent}, {Lupton},
  {Rockosi}, {Evans}, {Waddell}, {Anderson}, {Annis}, {Barentine}, {Bartoszek},
  {Bastian}, {Bracker}, {Brewington}, {Briegel}, {Brinkmann}, {Brown}, {Carr},
  {Czarapata}, {Drennan}, {Dombeck}, {Federwitz}, {Gillespie}, {Gonzales},
  {Hansen}, {Harvanek}, {Hayes}, {Jordan}, {Kinney}, {Klaene}, {Kleinman},
  {Kron}, {Kresinski}, {Lee}, {Limmongkol}, {Lindenmeyer}, {Long}, {Loomis},
  {McGehee}, {Mantsch}, {Neilsen}, {Neswold}, {Newman}, {Nitta}, {Peoples},
  {Pier}, {Prieto}, {Prosapio}, {Rivetta}, {Schneider}, {Snedden}, \&
  {Wang}}]{gunn06}
{Gunn} J.~E. {et~al.}, 2006, \aj, 131, 2332

\bibitem[{{Gustafsson} {et~al}\mbox{.}(2008){Gustafsson}, {Edvardsson},
  {Eriksson}, {J{\o}rgensen}, {Nordlund}, \& {Plez}}]{gustafsson08}
{Gustafsson} B., {Edvardsson} B., {Eriksson} K., {J{\o}rgensen} U.~G.,
  {Nordlund} {\AA}., {Plez} B., 2008, \aap, 486, 951

\bibitem[Han et al.(2014)]{han14} Han, E., Wang, S.~X., 
Wright, J.~T., et al.\ 2014, \pasp, 126, 827

\bibitem[{{Hartkopf}, {Mason} \& {Worley}(2001){Hartkopf}, {Mason}, \& {Worley}}]{hartkopf01} 
{Hartkopf} W.~I., {Mason} B.~D., {Worley} C.~E., 2001, \aj, 122, 3472

\bibitem[Hartkopf \& Mason(2009)]{hartkopf09} Hartkopf, W.~I., \& Mason, B.~D.\ 2009, \aj, 138, 813 

\bibitem[Hartmann et al.(1998)]{hartmann98} Hartmann, L., Calvet, N., Gullbring, E., \& D'Alessio, P.\ 1998, \apj, 495, 385

\bibitem[{{Hatzes}(2013)}]{hatzes13}
{Hatzes} A.~P., 2013, \apj, 770, 133

\bibitem[Helled et al.(2014)]{helled14} Helled, R., Bodenheimer, 
P., Podolak, M., et al.\ 2014, Protostars and Planets VI, 643

\bibitem[{{Henry}(1999)}]{henry99}
{Henry} G.~W., 1999, \pasp, 111, 845

\bibitem[{{Horch} {et~al}\mbox{.}(2010){Horch}, {Falta}, {Anderson}, {DeSousa},
  {Miniter}, {Ahmed}, \& {van Altena}}]{horch10}
{Horch} E.~P., {Falta} D., {Anderson} L.~M., {DeSousa} M.~D., {Miniter} C.~M.,
  {Ahmed} T., {van Altena} W.~F., 2010, \aj, 139, 205

\bibitem[{{Horch} {et~al}\mbox{.}(2008){Horch}, {van Altena}, {Cyr},
  {Kinsman-Smith}, {Srivastava}, \& {Zhou}}]{horch08}
{Horch} E.~P., {van Altena} W.~F., {Cyr}, Jr. W.~M., {Kinsman-Smith} L.,
  {Srivastava} A., {Zhou} J., 2008, \aj, 136, 312

\bibitem[Howard et al.(2010)]{ho10} Howard, A.~W., Johnson, J.~A., Marcy, G.~W., et al.\ 2010, \apj, 721, 1467 


\bibitem[{{Howard} {et~al}\mbox{.}(2014){Howard}, {Marcy}, {Fischer},
  {Isaacson}, {Muirhead}, {Henry}, {Boyajian}, {von Braun}, {Becker}, {Wright},
  \& {Johnson}}]{howard14}
{Howard} A.~W. {et~al.}, 2014, \apj, 794, 51

\bibitem[{{Howard} {et~al}\mbox{.}(2010){Howard}, {Marcy}, {Johnson},
  {Fischer}, {Wright}, {Isaacson}, {Valenti}, {Anderson}, {Lin}, \&
  {Ida}}]{howard10}
{Howard} A.~W. {et~al.}, 2010, Science, 330, 653

\bibitem[\protect\citeauthoryear{Kane 
\& von Braun}{2008}]{kane08} Kane S.~R., von Braun K., 2008, ApJ, 689, 492 

\bibitem[{{Konacki}(2005)}]{Konacki05}
{Konacki} M., 2005, \nat, 436, 230

\bibitem[{{Kratter} \& {Murray-Clay}(2011)}]{kratter11}
{Kratter} K.~M., {Murray-Clay} R.~A., 2011, \apj, 740, 1

\bibitem[{{Kraus} {et~al}\mbox{.}(2012){Kraus}, {Ireland}, {Hillenbrand}, \&
  {Martinache}}]{kraus12}
{Kraus} A.~L., {Ireland} M.~J., {Hillenbrand} L.~A., {Martinache} F., 2012,
  \apj, 745, 19

\bibitem[{{Kupka} {et~al}\mbox{.}(1999){Kupka}, {Piskunov}, {Ryabchikova},
  {Stempels}, \& {Weiss}}]{kupka99}
{Kupka} F., {Piskunov} N., {Ryabchikova} T.~A., {Stempels} H.~C., {Weiss}
  W.~W., 1999, \aaps, 138, 119

\bibitem[{{Lagrange} {et~al}\mbox{.}(2006){Lagrange}, {Beust}, {Udry},
  {Chauvin}, \& {Mayor}}]{lagrange06}
{Lagrange} A.-M., {Beust} H., {Udry} S., {Chauvin} G., {Mayor} M., 2006, \aap,
  459, 955

\bibitem[{{Lovis} {et~al}\mbox{.}(2006){Lovis}, {Mayor}, {Pepe}, {Alibert},
  {Benz}, {Bouchy}, {Correia}, {Laskar}, {Mordasini}, {Queloz}, {Santos},
  {Udry}, {Bertaux}, \& {Sivan}}]{lovis06}
{Lovis} C. {et~al.}, 2006, \nat, 441, 305

\bibitem[\protect\citeauthoryear{Ma et al.}{2013}]{ma13} Ma B., et al., 2013, AJ, 145, 20

\bibitem[{{Ma} \& {Ge}(2014)}]{ma14}
{Ma} B., {Ge} J., 2014, \mnras, 439, 2781

\bibitem[{{Mahadevan} {et~al}\mbox{.}(2008){Mahadevan}, {van Eyken}, {Ge},
  {DeWitt}, {Fleming}, {Cohen}, {Crepp}, \& {Vanden Heuvel}}]{mahadevan08}
{Mahadevan} S., {van Eyken} J., {Ge} J., {DeWitt} C., {Fleming} S.~W., {Cohen}
  R., {Crepp} J., {Vanden Heuvel} A., 2008, \apj, 678, 1505

\bibitem[{{Marcy} \& {Butler}(1992)}]{marcy92}
{Marcy} G.~W., {Butler} R.~P., 1992, \pasp, 104, 270

\bibitem[{{Mayer} {et~al}\mbox{.}(2005){Mayer}, {Wadsley}, {Quinn}, \&
  {Stadel}}]{mayer05}
{Mayer} L., {Wadsley} J., {Quinn} T., {Stadel} J., 2005, \mnras, 363, 641

\bibitem[Mayor \& Queloz(1995)]{mayor95} Mayor, M., \& Queloz, D.\ 1995, \nat, 378, 355 

\bibitem[Marzari et al.(2005)]{marzari05} Marzari, F., Weidenschilling, S.~J., Barbieri, M., \& Granata, V.\ 2005, \apj, 618, 502

\bibitem[{{McArthur} {et~al}\mbox{.}(2004){McArthur}, {Endl}, {Cochran},
  {Benedict}, {Fischer}, {Marcy}, {Butler}, {Naef}, {Mayor}, {Queloz}, {Udry},
  \& {Harrison}}]{mcArthur04}
{McArthur} B.~E. {et~al.}, 2004, \apjl, 614, L81

\bibitem[{{Mugrauer} \& {Neuh{\"a}user}(2005)}]{mugrauer05}
{Mugrauer} M., {Neuh{\"a}user} R., 2005, \mnras, 361, L15

\bibitem[{{Muterspaugh} {et~al}\mbox{.}(2010){Muterspaugh}, {Lane}, {Kulkarni},
  {Konacki}, {Burke}, {Colavita}, {Shao}, {Hartkopf}, {Boss}, \&
  {Williamson}}]{muterspaugh10}
{Muterspaugh} M.~W. {et~al.}, 2010, \aj, 140, 1657

\bibitem[Nakajima et al.(1995)]{nakajima95} Nakajima, T., Oppenheimer, B.~R., Kulkarni, S.~R., et al.\ 1995, \nat, 378, 463 

\bibitem[{{Nelson}(2000)}]{nelson00b}
{Nelson} A.~F., 2000, \apjl, 537, L65

\bibitem[{{Nordstrom} {et~al}\mbox{.}(2008){Nordstrom}, {Mayor}, {Andersen},
  {Holmberg}, {Pont}, {Jorgensen}, {Olsen}, {Udry}, \& {Mowlavi}}]{Nordstrom08}
{Nordstrom} B. {et~al.}, 2008, VizieR Online Data Catalog, 5117, 0

\bibitem[Ortiz et al.(2016)]{ortiz16} Ortiz, M., Reffert, S., Trifonov, T., et al.\ 2016, arXiv:1608.00963 

\bibitem[Oscoz et al.(2008)]{oscoz08} Oscoz, A., Rebolo, R., 
L{\'o}pez, R., et al.\ 2008, \procspie, 7014, 701447

\bibitem[Paegert et al.(2015)]{pag15} Paegert, M., Stassun, 
K.~G., De Lee, N., et al.\ 2015, \aj, 149, 186

\bibitem[{{Perryman}(1997)}]{perryman97}
{Perryman} M.~A.~C., 1997, in ESA Special Publication, Vol. 402, Hipparcos -
  Venice '97, {Bonnet} R.~M., {H{\o}g} E., {Bernacca} P.~L., {Emiliani} L.,
  {Blaauw} A., {Turon} C., {Kovalevsky} J., {Lindegren} L., {Hassan} H.,
  {Bouffard} M., {Strim} B., {Heger} D., {Perryman} M.~A.~C., {Woltjer} L.,
  eds., pp. 1--4

\bibitem[{{Plez} \& {Cohen}(2005)}]{plez05}
{Plez} B., {Cohen} J.~G., 2005, \aap, 434, 1117

\bibitem[{{Queloz} {et~al}\mbox{.}(2001){Queloz}, {Henry}, {Sivan}, {Baliunas},
  {Beuzit}, {Donahue}, {Mayor}, {Naef}, {Perrier}, \& {Udry}}]{queloz01}
{Queloz} D. {et~al.}, 2001, \aap, 379, 279

\bibitem[{{Queloz} {et~al}\mbox{.}(2000){Queloz}, {Mayor}, {Weber},
  {Bl{\'e}cha}, {Burnet}, {Confino}, {Naef}, {Pepe}, {Santos}, \&
  {Udry}}]{queloz00}
{Queloz} D. {et~al.}, 2000, \aap, 354, 99

\bibitem[\protect\citeauthoryear{Rajpaul, Aigrain, 
\& Roberts}{2016}]{raj16} Rajpaul V., Aigrain S., Roberts S., 2016, MNRAS, 456, L6

\bibitem[{{Ramsey} {et~al}\mbox{.}(1998){Ramsey}, {Adams}, {Barnes}, {Booth},
  {Cornell}, {Fowler}, {Gaffney}, {Glaspey}, {Good}, {Hill}, {Kelton},
  {Krabbendam}, {Long}, {MacQueen}, {Ray}, {Ricklefs}, {Sage}, {Sebring},
  {Spiesman}, \& {Steiner}}]{ramsey98}
{Ramsey} L.~W. {et~al.}, 1998, in Society of Photo-Optical Instrumentation
  Engineers (SPIE) Conference Series, Vol. 3352, Advanced Technology Optical/IR
  Telescopes VI, {Stepp} L.~M., ed., pp. 34--42

\bibitem[Rebolo et al.(1995)]{rebolo95} Rebolo, R., Zapatero Osorio, M.~R., \& Mart{\'{\i}}n, E.~L.\ 1995, \nat, 377, 129

\bibitem[{{Rivera} {et~al}\mbox{.}(2005){Rivera}, {Lissauer}, {Butler},
  {Marcy}, {Vogt}, {Fischer}, {Brown}, {Laughlin}, \& {Henry}}]{rivera05}
{Rivera} E.~J. {et~al.}, 2005, \apj, 634, 625

\bibitem[{{Robinson} {et~al}\mbox{.}(2006){Robinson}, {Strader}, {Ammons},
  {Laughlin}, \& {Fischer}}]{Robinson06}
{Robinson} S.~E., {Strader} J., {Ammons} S.~M., {Laughlin} G., {Fischer} D.,
  2006, \apj, 637, 1102

\bibitem[{{Santos} {et~al}\mbox{.}(2004){Santos}, {Bouchy}, {Mayor}, {Pepe},
  {Queloz}, {Udry}, {Lovis}, {Bazot}, {Benz}, {Bertaux}, {Lo Curto},
  {Delfosse}, {Mordasini}, {Naef}, {Sivan}, \& {Vauclair}}]{santos04}
{Santos} N.~C. {et~al.}, 2004, \aap, 426, L19

\bibitem[{{Santos} {et~al}\mbox{.}(2002){Santos}, {Mayor}, {Naef}, {Pepe},
  {Queloz}, {Udry}, {Burnet}, {Clausen}, {Helt}, {Olsen}, \&
  {Pritchard}}]{santos02}
{Santos} N.~C. {et~al.}, 2002, \aap, 392, 215

\bibitem[{{Shetrone} {et~al}\mbox{.}(2007){Shetrone}, {Cornell}, {Fowler},
  {Gaffney}, {Laws}, {Mader}, {Mason}, {Odewahn}, {Roman}, {Rostopchin},
  {Schneider}, {Umbarger}, \& {Westfall}}]{shetrone07}
{Shetrone} M. {et~al.}, 2007, \pasp, 119, 556

\bibitem[{{Spiegel}, {Burrows} \& {Milsom}(2011){Spiegel}, {Burrows}, \&
  {Milsom}}]{spiegel11}
{Spiegel} D.~S., {Burrows} A., {Milsom} J.~A., 2011, \apj, 727, 57

\bibitem[{{Thebault}(2011)}]{thebault11}
{Thebault} P., 2011, Celestial Mechanics and Dynamical Astronomy, 111, 29

\bibitem[{{Thebault} \& {Haghighipour}(2014)}]{thebault14}
{Thebault} P., {Haghighipour} N., 2014, ArXiv e-prints

\bibitem[Thomas et al.(2016)]{Thomas16} Thomas, N., Ge, J., Grieves, N., Li, R., \& Sithajan, S.\ 2016, \pasp, 128, 045003 

\bibitem[Tody(1993)]{tody93} Tody, D.\ 1993, Astronomical Data 
Analysis Software and Systems II, 52, 173

\bibitem[{{Torres}, {Andersen} \& {Gim{\'e}nez}(2010){Torres}, {Andersen}, \&
  {Gim{\'e}nez}}]{Torres2010}
{Torres} G., {Andersen} J., {Gim{\'e}nez} A., 2010, \aapr, 18, 67

\bibitem[{{Tull}(1998)}]{tull98}
{Tull} R.~G., 1998, in Society of Photo-Optical Instrumentation Engineers
  (SPIE) Conference Series, Vol. 3355, Optical Astronomical Instrumentation,
  {D'Odorico} S., ed., pp. 387--398

\bibitem[{{Udry} {et~al}\mbox{.}(2006){Udry}, {Mayor}, {Benz}, {Bertaux},
  {Bouchy}, {Lovis}, {Mordasini}, {Pepe}, {Queloz}, \& {Sivan}}]{udry06}
{Udry} S. {et~al.}, 2006, \aap, 447, 361

\bibitem[{{van Eyken}, {Ge} \& {Mahadevan}(2010){van Eyken}, {Ge}, \&
  {Mahadevan}}]{vaneyken10}
{van Eyken} J.~C., {Ge} J., {Mahadevan} S., 2010, \apjs, 189, 156

\bibitem[{{van Eyken} {et~al}\mbox{.}(2004){van Eyken}, {Ge}, {Mahadevan}, \&
  {DeWitt}}]{vaneyken04}
{van Eyken} J.~C., {Ge} J., {Mahadevan} S., {DeWitt} C., 2004, \apjl, 600, L79

\bibitem[{{Wan} {et~al}\mbox{.}(2006){Wan}, {Ge}, {Guo}, {Zhao}, {Hariharan},
  \& {McDavitt}}]{wan06}
{Wan} X., {Ge} J., {Guo} P., {Zhao} B., {Hariharan} L., {McDavitt} D., 2006, in
  Society of Photo-Optical Instrumentation Engineers (SPIE) Conference Series,
  Vol. 6269, Society of Photo-Optical Instrumentation Engineers (SPIE)
  Conference Series, p.~2

\bibitem[{{Wang} {et~al}\mbox{.}(2012{\natexlab{a}}){Wang}, {Ge}, {Wan}, {De
  Lee}, \& {Lee}}]{wang13}
{Wang} J., {Ge} J., {Wan} X., {De Lee} N., {Lee} B., 2012{\natexlab{a}}, \pasp,
  124, 1159

\bibitem[{{Wang} {et~al}\mbox{.}(2012{\natexlab{b}}){Wang}, {Ge}, {Wan}, {Lee},
  \& {De Lee}}]{wang12a}
{Wang} J., {Ge} J., {Wan} X., {Lee} B., {De Lee} N., 2012{\natexlab{b}}, \pasp,
  124, 598

\bibitem[Wolszczan \& Frail(1992)]{zan92} Wolszczan, A., \& Frail, D.~A.\ 1992, \nat, 355, 145 


\bibitem[{{Wright}(2005)}]{wright05}{Wright} J.~T., 2005, \pasp, 117, 657

\bibitem[Wu \& Murray(2003)]{wu03} Wu, Y., \& Murray, N.\ 2003, \apj, 589, 605 

\bibitem[{{Xie} \& {Zhou}(2008)}]{xie08}
{Xie} J.-W., {Zhou} J.-L., 2008, \apj, 686, 570

\bibitem[{{Xie} \& {Zhou}(2009)}]{xie09}
{Xie} J.-W., {Zhou} J.-L., 2009, \apj, 698, 2066

\bibitem[{{Xie}, {Zhou} \& {Ge}(2010){Xie}, {Zhou}, \& {Ge}}]{xie10}
{Xie} J.-W., {Zhou} J.-L., {Ge} J., 2010, \apj, 708, 1566

\bibitem[{{Zhao} \& {Ge}(2006)}]{zhao06}
{Zhao} B., {Ge} J., 2006, in Society of Photo-Optical Instrumentation Engineers
  (SPIE) Conference Series, Vol. 6269, Society of Photo-Optical Instrumentation
  Engineers (SPIE) Conference Series, p.~2

\bibitem[{{Zucker} {et~al}\mbox{.}(2003){Zucker}, {Mazeh}, {Santos}, {Udry}, \&
  {Mayor}}]{zucker03}
{Zucker} S., {Mazeh} T., {Santos} N.~C., {Udry} S., {Mayor} M., 2003, \aap,
  404, 775

\bibitem[{{Zucker} {et~al}\mbox{.}(2004){Zucker}, {Mazeh}, {Santos}, {Udry}, \&
  {Mayor}}]{zucker04}
{Zucker} S., {Mazeh} T., {Santos} N.~C., {Udry} S., {Mayor} M., 2004, \aap,
  426, 695


\end{thebibliography}
\end{document}